\documentclass[twocolumn]{aastex63}

\pdfoutput=1 %for arXiv submission
\usepackage[figure,figure*]{hypcap}
\usepackage{footnote}
\usepackage{hyperref}
\usepackage{subfigure}
\usepackage{amsmath}
\usepackage{graphicx}

\usepackage{epsfig}
\usepackage{natbib}

\shorttitle{3-D CMZ II}
\shortauthors{Battersby et al.}

\begin{document}

\def\Msun{\hbox{M$_{\odot}$}}
\def\Lsun{\hbox{L$_{\odot}$}}
\def\kms{km~s$^{\rm -1}$}
\def\hcop{HCO$^{+}$}
\def\n2hp{N$_{2}$H$^{+}$}
\def\micron{$\mu$m}
\def\13CO{$^{13}$CO}
\def\etamb{$\eta_{\rm mb}$}
\def\Inu{I$_{\nu}$}
\def\kapnu{$\kappa _{\nu}$}
\def\ffore{f$_{\rm{fore}}$}
\def\tastar{T$_{A}^{*}$}
\def\nh3{NH$_{3}$}
\def\deg{$^{\circ}$}
\def\arcsec{$^{\prime\prime}$}
\def\arcmin{$^{\prime}$}
\def\Vlsr{\hbox{V$_{LSR}$}}
\def\newv2{\textcolor{orange}}

\title{3-D CMZ II: Hierarchical Structure Analysis of the Central Molecular Zone}

\author[0000-0002-6073-9320]{Cara Battersby}
\affiliation{University of Connecticut, Department of Physics, 196A Hillside Road, Unit 3046
Storrs, CT 06269-3046, USA}
\affiliation{Center for Astrophysics $|$ Harvard \& Smithsonian, MS-78, 60 Garden St., Cambridge, MA 02138 USA}

\author[0000-0001-7330-8856]{Daniel L. Walker}
\affiliation{UK ALMA Regional Centre Node, Jodrell Bank Centre for Astrophysics, Oxford Road, The University of Manchester, Manchester M13 9PL, United Kingdom}
\affiliation{University of Connecticut, Department of Physics, 196A Hillside Road, Unit 3046
Storrs, CT 06269-3046, USA}

\author[0000-0003-0410-4504]{Ashley Barnes}
\affiliation{European Southern Observatory (ESO), Karl-Schwarzschild-Straße 2, 85748 Garching, Germany}

\author[0000-0001-6431-9633]{Adam Ginsburg}
\affiliation{University of Florida Department of Astronomy, Bryant Space Science Center, Gainesville, FL, 32611, USA}

\author[0000-0002-5776-9473]{Dani Lipman}
\affiliation{University of Connecticut, Department of Physics, 196A Hillside Road, Unit 3046
Storrs, CT 06269-3046, USA}

\author[0009-0005-9578-2192]{Danya Alboslani}
\affiliation{University of Connecticut, Department of Physics, 196A Hillside Road, Unit 3046
Storrs, CT 06269-3046, USA}

\author[0000-0003-0946-4365]{H Perry Hatchfield}
\affiliation{{Jet Propulsion Laboratory, California Institute of Technology, 4800 Oak Grove Drive, Pasadena, CA, 91109, USA}}
\affiliation{University of Connecticut, Department of Physics, 196A Hillside Road, Unit 3046
Storrs, CT 06269-3046, USA}

\author[0000-0001-8135-6612]{John Bally}
\affiliation{CASA, University of Colorado, 389-UCB, Boulder, CO 80309}

\author[0000-0001-6708-1317]{Simon C.~O.\ Glover}
\affiliation{Universit\"{a}t Heidelberg, Zentrum f\"{u}r Astronomie, Institut f\"{u}r Theoretische Astrophysik, Albert-Ueberle-Str.\ 2, 69120 Heidelberg, Germany}

\author[0000-0001-9656-7682]{Jonathan~D.~Henshaw}
\affiliation{Astrophysics Research Institute, Liverpool John Moores University, 146 Brownlow Hill, Liverpool L3 5RF, UK}
\affiliation{Max-Planck-Institut f\"ur Astronomie, K\"onigstuhl 17, D-69117 Heidelberg, Germany}

\author[0000-0003-4140-5138]{Katharina Immer}
\affiliation{European Southern Observatory (ESO), Karl-Schwarzschild-Straße 2, 85748 Garching, Germany}

\author[0000-0002-0560-3172]{Ralf S.\ Klessen}
\affiliation{Universit\"{a}t Heidelberg, Zentrum f\"{u}r Astronomie, Institut f\"{u}r Theoretische Astrophysik, Albert-Ueberle-Str.\ 2, 69120 Heidelberg, Germany}
\affiliation{Universit\"{a}t Heidelberg, Interdisziplin\"{a}res Zentrum f\"{u}r Wissenschaftliches Rechnen, Im Neuenheimer Feld 225, 69120 Heidelberg, Germany}
\affiliation{Center for Astrophysics $|$ Harvard \& Smithsonian, MS-78, 60 Garden St., Cambridge, MA 02138 USA}
\affiliation{Radcliffe Institute for Advanced Studies at Harvard University, 10 Garden Street, Cambridge, MA 02138, USA}

\author[0000-0001-6353-0170]{Steven N.~Longmore}
\affiliation{Astrophysics Research Institute, Liverpool John Moores University, 146 Brownlow Hill, Liverpool L3 5RF, UK}
\affiliation{Cosmic Origins Of Life (COOL) Research DAO, coolresearch.io}

\
\author[0000-0001-8782-1992]{Elisabeth A.~C.~Mills}
\affiliation{Department of Physics and Astronomy, University of Kansas, 1251 Wescoe Hall Drive, Lawrence, KS 66045, USA}

\author{Sergio Molinari}
\affiliation{INAF - Istituto di Astrofisica e Planetologia Spaziale, Via Fosso del Cavaliere 100, I-00133 Roma, Italy}

\author[0000-0002-0820-1814]{Rowan Smith}
\affiliation{Scottish Universities Physics Alliance (SUPA), School of Physics and Astronomy, University of St Andrews, North Haugh, St Andrews KY16 9SS, UK}
\affiliation{Jodrell Bank Centre for Astrophysics, Department of Physics and Astronomy, University of Manchester, Oxford Road, Manchester M13 9PL, UK}

\author[0000-0001-6113-6241]{Mattia C.\ Sormani}
\affiliation{Universit{\`a} dell’Insubria, via Valleggio 11, 22100 Como, Italy}

\author[0000-0002-9483-7164]{Robin~G.~Tress}
\affiliation{Institute of Physics, Laboratory for Galaxy Evolution and Spectral Modelling, EPFL, Observatoire de Sauverny, Chemin Pegasi 51, 1290 Versoix, Switzerland}

\author[0000-0003-2384-6589]{Qizhou Zhang}
\affiliation{Center for Astrophysics $|$ Harvard \& Smithsonian, MS-78, 60 Garden St., Cambridge, MA 02138 USA}

\begin{abstract}
The Central Molecular Zone (CMZ) is the way station at the heart of our Milky Way Galaxy, connecting gas flowing in from Galactic scales with the central nucleus.  
Key open questions remain about its 3-D structure, star formation properties, and role in regulating this gas inflow.
In this work, we identify a hierarchy of discrete structures in the CMZ using column density maps from Paper I (Battersby et al., submitted).
We calculate the physical ($N$(H$_2$), $T_{\rm{dust}}$, mass, radius) and kinematic (HNCO, HCN, and HC$_3$N moments) properties of each structure as well as their bolometric luminosities and star formation rates (SFRs). 
We compare these properties with regions in the Milky Way disk and external galaxies.
Despite the fact that the CMZ overall is well below the Gao-Solomon dense gas star-formation relation (and in modest agreement with the Schmidt-Kennicutt relation), individual structures on the scale of molecular clouds generally follow these star-formation relations and agree well with other Milky Way and extragalactic regions.
We find that individual CMZ structures require a large external pressure ($P_e$/k$_B$ $> 10^{7-9}$ K cm$^{-3}$) to be considered bound, however simple estimates suggest that most CMZ molecular-cloud-sized structures are consistent with being in pressure-bounded virial equilibriuim. 
We perform power-law fits to the column density probability distribution functions (N-PDFs) of the inner 100 pc, SgrB2, and the outer 100 pc of the CMZ as well as several individual molecular cloud structures and find generally steeper power-law slopes ($-9<\alpha<-2$) compared with the literature ($-6 < \alpha < -1$). 
\end{abstract}

\section{Introduction}
\label{sec:intro}

The inner few hundred parsecs of the Milky Way, known as the ``Central Molecular Zone" (CMZ), contains the largest reservoir of dense molecular gas in the Galaxy \citep{Morris1996,Henshaw2023}. 
The interstellar medium (ISM) in the CMZ is extreme in many ways compared with that of the Galactic disk and may have more similarities with local starburst systems and high-$z$ ($z\sim 2$) galaxies than with regions in the disk \citep{Kruijssen2013}.
For example, gas in the CMZ has higher densities \citep[e.g.][]{Mills2018b, Schmiedeke2016, Walker2016, Bally1987, Gusten1983} and temperatures \citep[e.g.][]{Krieger2017, Ginsburg2016, Mills2013} than gas in the Galactic disk and is also more turbulent \citep[e.g.][]{Krieger2020, Kauffmann2017a, Federrath2016, Henshaw2016a, Henshaw2016b, Shetty2012, Bally1987} with stronger magnetic fields \citep[e.g.][]{Crutcher1996, Pillai2015}. The CMZ is also subject to a stronger UV background field \citep[e.g.][]{Clark2013, Goicoechea2004, Lis2001}, higher cosmic ray ionization rates \citep[e.g.][]{Padovani2020, Harada2015, Indriolo2015, Goto2013, Oka2005}, frequent X-ray flares \citep{Terrier2010, Terrier2018, Clavel2013}, as well as unique dynamical stresses \citep[e.g.][]{Tress2020, Sormani2020, Kruijssen2019, Sormani2018b, Krumholz2017, Longmore2013b}. \citet{Kruijssen2013} argue that these locally `extreme' properties share many similarities with high-redshift galaxies, making the CMZ ideally suited to understand the physics in these distant galaxies up close at a distance of only 8.2~kpc \citep{Reid2019,Gravity19, Gravity2021}.

Gas flows into the CMZ from the Galactic disk along the bar dust lanes with an estimated inflow rate of 0.8 $\pm$ 0.6 \Msun yr$^{-1}$ \citep[e.g.][]{Hatchfield2021, Sormani2019}. Various measurements indicate that the CMZ's star formation rate (SFR) has been roughly constant over the last 5 Myr at a value of about 0.07 \Msun  yr$^{-1}$ \citep[][and references within]{Barnes2017, Henshaw2023, Hatchfield2024}. The gas inflow is much greater than the SFR, which highlights the importance of the CMZ's role in regulating gas inflow to the central nucleus. The CMZ engages in prolific star formation in the SgrB2 molecular cloud \citep[e.g.][]{Meng2022, Schworer2019, Ginsburg2018}, but overall has a lower than expected SFR \citep{Longmore2012, Immer2012}. While the CMZ is roughly consistent with the Schmidt-Kennicutt relation between gas and SFR, it falls about an order of magnitude below the Gao-Solomon dense gas star formation relation (discussed in detail in \citealt{Henshaw2023} and references within). 
The 3-D structure of this region is key to understanding why this region is `under-performing' in star formation and its role in regulating gas inflow from the Galactic disk to the central nucleus. \textit{Without an accurate 3-D model, we cannot trace gas flows through the CMZ onto the central nucleus, nor can we understand the patterns and causes of its star formation.}

Many works have studied the CMZ overall \citep[e.g.][]{Ginsburg2016, Krieger2017, Kauffmann2017a, Kauffmann2017b, Hatchfield2020, Lu2021} or individual regions within \citep[e.g.][]{Kendrew2013, Lu2017, Walker2018, Walker2021}. In this work, our aim is to create a comprehensive, hierarchical catalog of structures in the CMZ from the entire CMZ to individual molecular clouds. We apply simple and common analyses that allow us to inter-compare this unique environment with many regions in the literature, from the disk of the Milky Way to more distant galaxies. 

This paper is the second in a series focused on understanding the 3-D structure of our CMZ. Battersby et al. (submitted, hereafter Paper I) presents the column density and temperature measurements (derived from modified blackbody fits to Herschel data) that provide the basis of our hierarchical structure catalog. Walker et al. (submitted, hereafter Paper III) performs a comprehensive kinematic analysis of molecular clouds in the catalog and uses molecular line absorption to provide evidence for whether clouds are on the near or far side of the CMZ. Lipman et al. (submitted, hereafter Paper IV) uses Spitzer 8\micron~and Herschel 70\micron~dust extinction maps to constrain the 3-D position of clouds in the CMZ and compare with existing models.

This paper is organized as follows. In Section \ref{sec:methods} we describe the data used in this work, and the segmentation of the column density map into a hierarchicial structure catalog. Section \ref{sec:analysis} presents the physical and kinematic properties of each structure in this catalog as well as their luminosities and SFRs. This section also presents the method to fit power-laws to the column density probability distribution functions (N-PDFs) of select structures in the catalog. 
The properties derived (sizes, linewidths, surface densities, and SFRs) are compared with other Milky Way and extragalactic regions in Section \ref{sec:discussion}. In Section \ref{sec:discussion} we also plot our power-law slopes against our measured SFRs. We summarize this work and conclude in Section \ref{sec:conclusions}.

%now made in dendro_cmz_higal_v3.ipynb
% (previously: NEW_dendro_cmz_higal.ipynb) and figure_combination_higal_cmz.pptx" in PAPER/ALL_figures/ folder
\begin{figure*}
\begin{centering}
\includegraphics[trim=0 15mm 0 0, clip, width=1\textwidth]{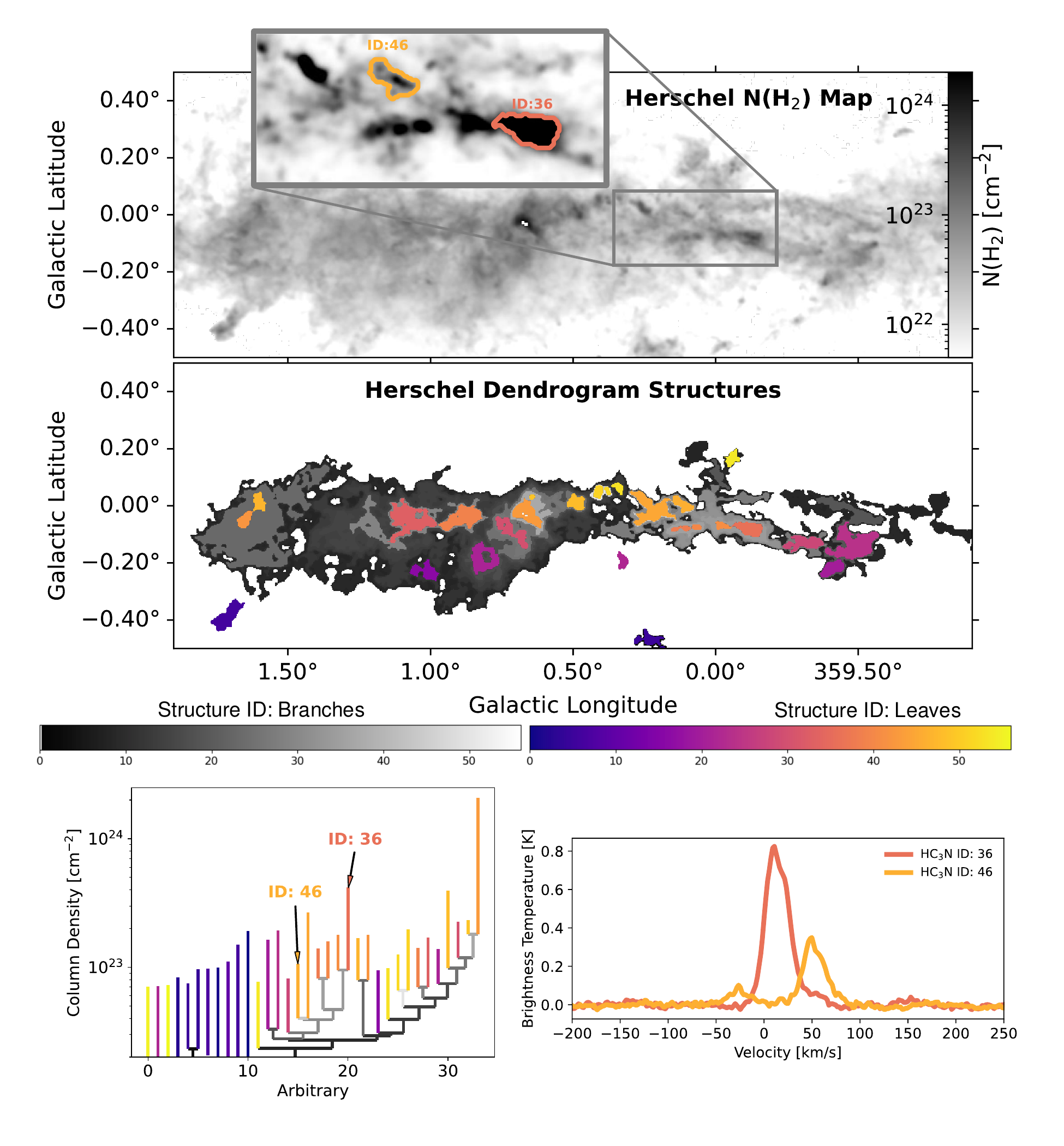}
\caption{A dendrogram segmentation algorithm is used to create a hierarchical structure catalog from our Herschel column density map (top panel), segmenting the CMZ into its largest scales (structure ID: 9 is the entire CMZ) with a complete hierarchy to individual cloud scales. The middle shows the dendrogram segmentation as a map of structures colored by their IDs, with branches colored in a grayscale colorbar and leaves in a plasma colorbar. The lower left shows the associated dendrogram tree with the same structure ID color scheme. The zoom-in at the very top highlights two individual clouds in the Herschel column density map, labeled as structures 46 and 36 (the `dust ridge bridge' and 20 \kms~cloud respectively). The location of each of these is highlighted in the dendrogram tree on the bottom left and the associated integrated spectra for these structures in the HC$_3$N line are shown in the bottom right panel.}
\end{centering}
\label{fig:dendro}
\end{figure*}

\section{Methods}
\label{sec:methods}

%For all of our analyses we assume a distance to the CMZ of 8.2 kpc \citep{Reid2019, Gravity19, Gravity2021}.

\subsection{Data}
\label{sec:obs}

In Paper I, we present column density and dust temperature maps of the inner 40\deg~of the Galaxy. These maps were derived by performing modified blackbody fits to Herschel data from 160 to 500 \micron~from the Hi-Gal survey \citep{Molinari2011}. We smoothed the data to the lowest common resolution of 36\arcsec~\citep{Molinari2016} or about 1.4 pc at the Galactic Center. Using an automated iterative approach, we identified and subtracted the contribution to the Herschel maps from diffuse cirrus emission in the fore/background of the dense structure in our maps. The maps used in this work have been fore/background subtracted and are our best estimate of the dense gas column density and dust temperature in the CMZ on 1 pc scales.

In addition to the Herschel data from Paper I, we take advantage of additional extant data for our analysis. This includes mid-infrared data from the Spitzer space telescope, taken as part of the GLIMPSE \citep{Benjamin2003} and MIPSGAL \citep{Carey2009} surveys. These data are combined with the Herschel data in a companion paper \citep{Barnes2017} to produce bolometric luminosities. We use these luminosities to derive star formation rates (SFRs) reported in Table \ref{tab:luminosity_properties} as described in Section \ref{sec:sfr}. 

We also utilize data from the Mopra 22~m radio telescope of the Australia Telescope National Facility survey of the CMZ, the full observational and data reduction details of which were published by \citet{Jones2012}. In short, Mopra achieved a 39\arcsec~(1.5~pc at the Galactic Center) beam centered on frequencies from 85.3 -- 93.3~GHz with an 8~GHz bandpass and $\sim$3.6~km s$^{-1}$ spectral resolution, sampled in 1.8~\kms channels. A full kinematic decomposition of the CMZ using the HNCO 4(0,4)--3(0,3) spectral line data is available in \citet{Henshaw2016a}. We use these data to derive approximate kinematic properties of our catalog structures, reported in Table \ref{tab:kinematic_properties} and described in Section \ref{sec:kinematic_properties}.

\subsection{Categorizing the Hierarchical Structure with Dendrograms}
\label{sec:dendrograms}

A dendrogram is a tree-like diagram that can be used to represent the hierarchical structure of a given data-set by segmenting the structure using nested iso-density contours. This technique is used to divide the hierarchical structure in a map into discrete iso-density contours of contiguous and significant emission, and retains information on each structure from largest to smallest scales relevant for a given dataset. The use of dendrograms in astronomy is explained in more detail in \citet{Rosolowsky2008b}, \citet{Goodman2009}, or \citet{Shetty2012} for example. The lowest-level emission forms the `trunk' of the tree, followed by the `branches' and `leaves.' Dendrograms are particularly well-suited for regions of complex, hierarchical emission with structures spanning over an order of magnitude in spatial scale. For this reason, we have selected this method to identify structures in the CMZ using our 2-D map of column density from Herschel, which is sensitive from scales of 1-100 pc. 

We use the python package {\sc astrodendro}\footnote{\href{https://dendrograms.readthedocs.io/en/stable/}{https://dendrograms.readthedocs.io/en/stable/}} to segment our column density map into a hierarchical structure tree. The dendrogram structure tree and overall map of identified structures is shown in Figure \ref{fig:dendro}. Based on the known structures commonly referred to as clouds in the CMZ and through a detailed parameter study, we select dendrogram parameters of  [{\sc min\_value} = 2 $\times 10^{22}$ cm$^{-2}$, {\sc min\_delta} = 5 $\times 10^{22}$ cm$^{-2}$, {\sc min\_npix} = 10] for the decomposition. 
The minimum value chosen is about a factor of two larger than the typical background value in the CMZ of $N$(H$_2$)=10$^{22}$ cm$^{-2}$ while the step size is five times this. While the choice of parameters affects the size of individual regions, and therefore their physical properties, the full range of properties is robustly captured in the structure tree, regardless of the fine-tuning of these parameters. For the purpose of this and following works in this series, we tuned the dendrogram parameters to optimally capture well-known and long-studied CMZ cloud complexes.

Using the dendrogram parameters described above, we find 11 levels of hierarchical structure in the CMZ, with the highest column density leaf in SgrB2 Main. In total there are 57 structures (branches and leaves) identified. The dendrogram we derive is shown in Figure \ref{fig:dendro} and the properties of each structure are given in Table \ref{tab:general_properties}. The [$\ell$, $b$] diagram in Figure \ref{fig:dendro} shows the spatial extent of each dendrogram structure, colored by the object's structure ID from 0 to 56 based on the order in which structures are decomposed by the algorithm and without any physical significance. The corresponding dendrogram tree displayed in the lower-left of Figure \ref{fig:dendro} shows the full hierarchical tree as related to the peak column density of each structure on the y-axis. 
The `trunk' of the CMZ hierarchy is a single structure (ID: 9) -- a single, contiguous emission feature in ($\ell$, b) space, that is at the base of the large tree. The remaining structures outside this large tree in the lower-left panel of Figure \ref{fig:dendro} are isolated features in ($\ell$, b) space, not associated with the rest of the CMZ hierarchy.

\section{Analysis}
\label{sec:analysis}

\begin{table*}
\centering
\caption{General properties of the dendrogram structures. Shown for each structure is the source name, structure ID, corresponding leaf IDs in Paper III, corresponding region ID in the CMZoom survey, exact dendrogram area in parsec$^{2}$, central coordinates in degrees ($l, b$), median and peak column density (N$_{\textrm{H$_{2}$}}$), mass, radius (assuming the radius of a circle of equivalent area), volume density (assuming spherical symmetry), median and peak dust temperature (T$_{\textrm{d}}$), and colloquial name in the literature (where applicable). Structures are grouped by type (branch or leaf), and ordered within each group by descending area.}\resizebox{\linewidth}{!}{%
\begin{tabular}{ccccccccccccccc}
\hline
Name & ID & Paper III & CMZoom & Area & $l$ & $b$ & Median & Peak & Mass & Radius & Density & Median & Peak & Colloquial name\\
& & ID & region & & & & $N_{\textrm{H$_{2}$}}$ & $N_{\textrm{H$_{2}$}}$ & M & R& $n$ & T$_{\textrm{d}}$ & T$_{\textrm{d}}$ & / description\\ \hline
& & & & $\mathrm{pc^{2}}$ & $^{\circ}$ & $^{\circ}$ & $\mathrm{cm^{-2}}$ & $\mathrm{cm^{-2}}$ & $\mathrm{M_{\odot}}$ & pc & $\mathrm{cm^{-3}}$ & $\mathrm{K}$ & $\mathrm{K}$ &\\ \hline
Branches & & & & & & & & & & & &\\ \hline
G0.720-0.069 & 8 & - & - & 1.3E+04 & 0.720 & -0.069 & 3.5E+22 & 2.1E+24 & 1.5E+07 & 6.5E+01 & 1.9E+02 & 20 & 36 & -\footnote{Structure 8 is the entire CMZ along with an isolated peak} \\
G0.750-0.070 & 9 & - & - & 1.1E+04 & 0.750 & -0.070 & 3.9E+22 & 2.1E+24 & 1.3E+07 & 5.9E+01 & 2.2E+02 & 20 & 36 & Entire CMZ \\
G0.970-0.067 & 11 & - & - & 7.0E+03 & 0.970 & -0.067 & 4.6E+22 & 2.1E+24 & 1.0E+07 & 4.7E+01 & 3.3E+02 & 19 & 29 & Outer 100 pc $+$ SgrB2 $+$ dust ridge \\
G0.840-0.065 & 12 & - & - & 5.0E+03 & 0.840 & -0.065 & 5.2E+22 & 2.1E+24 & 7.7E+06 & 4.0E+01 & 4.2E+02 & 20 & 29 & -\footnote{Structures 12, 13, and 14 are all variations on the 1.1$^\circ$ complex $+$ SgrB2 $+$ dust ridge hierarchy with isolated peaks} \\
G0.830-0.062 & 13 & - & - & 4.6E+03 & 0.830 & -0.062 & 5.4E+22 & 2.1E+24 & 7.5E+06 & 3.8E+01 & 4.7E+02 & 20 & 29 & -$^{b}$ \\
G0.820-0.059 & 14 & - & - & 3.6E+03 & 0.820 & -0.059 & 6.3E+22 & 2.1E+24 & 6.6E+06 & 3.4E+01 & 5.8E+02 & 20 & 27 & -$^{b}$ \\
G0.800-0.060 & 15 & - & - & 2.6E+03 & 0.800 & -0.060 & 7.4E+22 & 2.1E+24 & 5.7E+06 & 2.9E+01 & 8.1E+02 & 20 & 27 & 1.1$^\circ$ complex $+$ SgrB2 \\
G359.900-0.071 & 19 & - & - & 2.0E+03 & -0.100 & -0.071 & 4.0E+22 & 4.2E+23 & 2.4E+06 & 2.5E+01 & 5.3E+02 & 25 & 36 & Inner 100pc - dust ridge \\
G0.020-0.058 & 27 & 8,20 & - & 1.4E+03 & 0.020 & -0.058 & 4.1E+22 & 4.2E+23 & 1.7E+06 & 2.1E+01 & 6.3E+02 & 25 & 36 & -\footnote{Structures 27, 31, and 32 are variations on the Inner 100 pc - dust ridge hierarchy. Structure 31 contains SgrA*, but its contribution to the average physical properties of this large region is negliglbe at far-ir wavelengths.} \\
G1.610-0.059 & 23 & - & - & 1.3E+03 & 1.610 & -0.059 & 4.1E+22 & 1.8E+23 & 1.5E+06 & 2.1E+01 & 5.6E+02 & 17 & 20 & 1.6$^\circ$ complex \\
G0.670-0.059 & 17 & - & - & 1.3E+03 & 0.670 & -0.059 & 9.4E+22 & 2.1E+24 & 3.7E+06 & 2.0E+01 & 1.6E+03 & 21 & 27 & Sgr B2 region \\
G0.040-0.054 & 31 & 12 & - & 9.4E+02 & 0.040 & -0.054 & 4.8E+22 & 4.2E+23 & 1.4E+06 & 1.7E+01 & 9.8E+02 & 25 & 36 & -$^{c}$ \\
G0.650-0.040 & 25 & - & - & 8.2E+02 & 0.650 & -0.040 & 1.1E+23 & 2.1E+24 & 3.0E+06 & 1.6E+01 & 2.5E+03 & 21 & 27 & -\footnote{Structures 25, 26, and 37 are all part of the Sgr B2 hierarchy} \\
G1.040-0.051 & 29 & - & - & 7.0E+02 & 1.040 & -0.051 & 7.4E+22 & 1.7E+23 & 1.3E+06 & 1.5E+01 & 1.3E+03 & 18 & 23 & 1.1$^\circ$ complex \\
G359.530-0.120 & 18 & 1,3,5 & - & 5.4E+02 & -0.470 & -0.120 & 4.1E+22 & 1.9E+23 & 5.6E+05 & 1.3E+01 & 8.8E+02 & 24 & 29 & SgrC $+$ far-side candidates \\
G0.670-0.038 & 26 & - & - & 4.8E+02 & 0.670 & -0.038 & 1.4E+23 & 2.1E+24 & 2.3E+06 & 1.2E+01 & 4.6E+03 & 21 & 26 & - \\
G359.960-0.078 & 32 & 11 & - & 4.4E+02 & -0.040 & -0.078 & 6.4E+22 & 4.2E+23 & 8.4E+05 & 1.2E+01 & 1.7E+03 & 24 & 35 & - \\
G0.660-0.023 & 37 & - & - & 2.2E+02 & 0.660 & -0.023 & 1.8E+23 & 2.1E+24 & 1.5E+06 & 8.4E+00 & 8.7E+03 & 21 & 25 & - \\
G0.210-0.003 & 44 & - & - & 2.2E+02 & 0.210 & -0.003 & 5.2E+22 & 2.7E+23 & 3.3E+05 & 8.4E+00 & 1.9E+03 & 25 & 28 & Dust Ridge Bridge $+$ Brick plus \\
G359.910-0.079 & 34 & - & - & 2.2E+02 & -0.090 & -0.079 & 9.0E+22 & 4.2E+23 & 5.5E+05 & 8.4E+00 & 3.2E+03 & 24 & 34 & 50 and 20 km/s Clouds \\
G0.220-0.490 & 3 & - & - & 1.2E+02 & 0.220 & -0.490 & 2.7E+22 & 9.7E+22 & 8.4E+04 & 6.1E+00 & 1.3E+03 & 16 & 25 & - \\
G0.110-0.079 & 35 & - & - & 8.8E+01 & 0.110 & -0.079 & 7.3E+22 & 1.6E+23 & 1.5E+05 & 5.3E+00 & 3.5E+03 & 23 & 25 & Three Little Pigs \\
G0.400+0.048 & 50 & - & - & 5.4E+01 & 0.400 & 0.048 & 7.6E+22 & 2.0E+23 & 1.0E+05 & 4.1E+00 & 5.0E+03 & 21 & 25 & Clouds d and c \\
\hline
Leaves & & & & & & & & & & & \\ \hline
G359.510-0.130 & 24 & 2 & 34,36\footnote[2]{Structures 20, 24, 28, and 45 are only partially covered by CMZoom} & 2.4E+02 & -0.490 & -0.130 & 4.9E+22 & 1.9E+23 & 2.9E+05 & 8.7E+00 & 1.5E+03 & 24 & 28 & SgrC \\
G1.070-0.049 & 33 & 29 & 5,6 & 2.2E+02 & 1.070 & -0.049 & 8.5E+22 & 1.7E+23 & 4.6E+05 & 8.4E+00 & 2.7E+03 & 18 & 23 & 1.1$^\circ$ Cloud East \\
G0.230-0.004 & 45 & 17 & 18,19$^{\dag}$ & 1.6E+02 & 0.230 & -0.004 & 5.4E+22 & 2.7E+23 & 2.5E+05 & 7.2E+00 & 2.3E+03 & 25 & 28 & Brick plus \\
G0.890-0.044 & 39 & 28 & 7 & 1.5E+02 & 0.890 & -0.044 & 9.1E+22 & 1.4E+23 & 3.1E+05 & 6.9E+00 & 3.3E+03 & 19 & 20 & 1.1$^\circ$ Cloud West \\
G0.820-0.190 & 21 & 27 & - & 1.3E+02 & 0.820 & -0.190 & 9.0E+22 & 1.4E+23 & 2.6E+05 & 6.4E+00 & 3.4E+03 & 19 & 20 & M0.8-0.2 ring \citep{Nonhebel2024} \\
G1.720-0.390 & 6 & - & - & 1.2E+02 & 1.720 & -0.390 & 3.8E+22 & 1.5E+23 & 1.2E+05 & 6.1E+00 & 1.8E+03 & 15 & 17 & - \\
G0.670-0.028 & 43 & 25 & 9 & 1.0E+02 & 0.670 & -0.028 & 3.6E+23 & 2.1E+24 & 1.2E+06 & 5.8E+00 & 2.1E+04 & 20 & 25 & Sgr B2 Main \\
G0.720-0.090 & 30 & 26 & 8 & 1.0E+02 & 0.720 & -0.090 & 1.4E+23 & 2.3E+23 & 3.2E+05 & 5.7E+00 & 6.0E+03 & 20 & 22 & SgrB2 `extended' \\
G359.690-0.130 & 28 & 6 & 31$^{\dag}$ & 9.8E+01 & -0.310 & -0.130 & 3.7E+22 & 8.2E+22 & 8.7E+04 & 5.6E+00 & 1.7E+03 & 24 & 26 & bridge from 20 km/s to SgrC \\
G359.880-0.081 & 36 & 9 & 28 & 7.8E+01 & -0.120 & -0.081 & 1.6E+23 & 4.2E+23 & 3.1E+05 & 5.0E+00 & 8.6E+03 & 21 & 26 & 20 km/s Cloud \\
G1.020-0.230 & 16 & - & - & 7.6E+01 & 1.020 & -0.230 & 3.6E+22 & 9.5E+22 & 6.5E+04 & 4.9E+00 & 1.9E+03 & 20 & 21 & - \\
G0.550-0.870 & 0 & - & - & 7.0E+01 & 0.550 & -0.870 & 3.3E+22 & 1.9E+23 & 7.1E+04 & 4.7E+00 & 2.4E+03 & 19 & 27 & - \\
G359.600-0.220 & 20 & 4 & 33$^{\dag}$ & 6.6E+01 & -0.400 & -0.220 & 4.8E+22 & 1.6E+23 & 8.1E+04 & 4.6E+00 & 2.9E+03 & 23 & 25 & peak south of SgrC \\
G0.240-0.470 & 5 & - & - & 5.8E+01 & 0.240 & -0.470 & 3.0E+22 & 7.5E+22 & 4.4E+04 & 4.3E+00 & 1.9E+03 & 16 & 25 & - \\
G0.490+0.008 & 48 & 23 & 11 & 5.4E+01 & 0.490 & 0.008 & 1.4E+23 & 3.9E+23 & 2.0E+05 & 4.1E+00 & 1.0E+04 & 20 & 22 & Clouds e and f \\
G0.120+0.003 & 46 & 15 & 22$^{\dag}$ & 5.0E+01 & 0.120 & 0.003 & 5.3E+22 & 1.1E+23 & 6.2E+04 & 4.0E+00 & 3.3E+03 & 25 & 28 & Dust ridge Bridge \\
G1.600+0.012 & 47 & 30 & 4 & 4.4E+01 & 1.600 & 0.012 & 9.7E+22 & 1.7E+23 & 1.0E+05 & 3.7E+00 & 6.8E+03 & 16 & 17 & 1.6$^{\circ}$ Cloud North \\
G1.650-0.052 & 42 & 31 & 3 & 4.4E+01 & 1.650 & -0.052 & 1.0E+23 & 1.8E+23 & 1.0E+05 & 3.7E+00 & 6.8E+03 & 15 & 16 & 1.6$^{\circ}$ Cloud South \\
G359.940+0.160 & 54 & - & - & 3.8E+01 & -0.060 & 0.160 & 3.2E+22 & 7.7E+22 & 3.2E+04 & 3.5E+00 & 2.6E+03 & 19 & 24 & - \\
G0.330-0.190 & 22 & 18 & 17 & 3.4E+01 & 0.330 & -0.190 & 2.8E+22 & 7.1E+22 & 2.5E+04 & 3.3E+00 & 2.4E+03 & 22 & 27 & - \\
G3.430-0.350 & 10 & - & - & 3.4E+01 & 3.430 & -0.350 & 3.4E+22 & 1.1E+23 & 2.9E+04 & 3.3E+00 & 2.8E+03 & 15 & 20 & - \\
G0.200-0.520 & 4 & - & - & 3.2E+01 & 0.200 & -0.520 & 3.0E+22 & 9.7E+22 & 2.7E+04 & 3.2E+00 & 2.8E+03 & 16 & 18 & - \\
G0.410+0.048 & 51 & 21 & 12 & 2.6E+01 & 0.410 & 0.048 & 9.7E+22 & 2.0E+23 & 6.2E+04 & 2.9E+00 & 8.8E+03 & 21 & 23 & Cloud d \\
G359.980-0.071 & 41 & 10 & 26 & 2.2E+01 & -0.020 & -0.071 & 1.2E+23 & 1.8E+23 & 6.2E+04 & 2.6E+00 & 1.2E+04 & 24 & 25 & 50 km/s Cloud \\
G0.340+0.060 & 53 & 19 & 15 & 2.0E+01 & 0.340 & 0.060 & 4.8E+22 & 9.9E+22 & 2.5E+04 & 2.5E+00 & 5.5E+03 & 23 & 26 & Cloud b \\
G0.120-0.081 & 38 & 14,16 & 20,21 & 2.0E+01 & 0.120 & -0.081 & 9.0E+22 & 1.4E+23 & 4.4E+04 & 2.5E+00 & 9.7E+03 & 22 & 23 & Straw and Sticks Clouds \\
G358.460-0.390 & 7 & - & - & 1.7E+01 & -1.540 & -0.390 & 3.3E+22 & 9.8E+22 & 1.5E+04 & 2.3E+00 & 4.3E+03 & 17 & 18 & - \\
G356.660+0.560 & 56 & - & - & 1.4E+01 & -3.340 & 0.560 & 2.6E+22 & 7.3E+22 & 9.1E+03 & 2.1E+00 & 3.4E+03 & 11 & 13 & - \\
G0.070-0.076 & 40 & 13 & 23 & 1.2E+01 & 0.070 & -0.076 & 1.1E+23 & 1.6E+23 & 3.0E+04 & 1.9E+00 & 1.5E+04 & 21 & 22 & Stone Cloud \\
G0.090-0.660 & 2 & - & - & 1.1E+01 & 0.090 & -0.660 & 3.0E+22 & 8.3E+22 & 8.7E+03 & 1.9E+00 & 4.4E+03 & 18 & 20 & - \\
G0.380+0.050 & 52 & 21 & 14 & 9.0E+00 & 0.380 & 0.050 & 8.7E+22 & 1.3E+23 & 1.8E+04 & 1.7E+00 & 1.3E+04 & 21 & 24 & Cloud c \\
G356.520+0.210 & 55 & - & - & 8.0E+00 & -3.480 & 0.210 & 2.4E+22 & 7.0E+22 & 5.2E+03 & 1.6E+00 & 4.4E+03 & 13 & 13 & - \\
G357.070-0.770 & 1 & - & - & 7.0E+00 & -2.930 & -0.770 & 2.9E+22 & 9.9E+22 & 5.0E+03 & 1.5E+00 & 5.1E+03 & 9 & 12 & - \\
G0.650+0.030 & 49 & 24 & - & 6.0E+00 & 0.650 & 0.030 & 2.0E+23 & 2.3E+23 & 2.8E+04 & 1.4E+00 & 3.5E+04 & 19 & 20 & small isolated peak in SgrB2 \\
\hline \hline
\end{tabular}%
\label{tab:general_properties}
}
\end{table*}

\subsection{Table Overview}
\label{sec:tables}
We report on the physical and kinematic properties of each dendrogram structure derived in this work in Tables \ref{tab:general_properties} and \ref{tab:kinematic_properties} and on their luminiosities in Table \ref{tab:luminosity_properties}.
Table \ref{tab:general_properties} reports on the name, structure ID, physical area, central coordinates, median and peak column density, mass, effective radius, median and peak dust temperature, as well as a colloquial name for each structure or a brief description. Table \ref{tab:kinematic_properties} reports on the moment 0, 1, and 2 results for the HNCO $4_{0,4}-3_{0,3}$, HCN $1-0$, and HC$_3$N $10-9$ transitions. Table \ref{tab:luminosity_properties} reports on luminosities and SFRs inferred from the Paper I modified blackbody fits.

The catalog presented here is for the full CMZ hierarchy. Paper III (Walker et al., submitted) is focused on the individual molecular clouds in the CMZ within this hierarchy. Therefore, in Paper III, we create a molecular cloud catalog using primarily the \textit{leaves} of the dendrogram structures from this work with a few small changes. Full details are described in Paper III, but the changes include: adding a few lower column-density molecular clouds to the catalog which are at critical positions in the CMZ orbit for determining near/far distances, splitting the Straw and Sticks molecular clouds (structure ID 38 in this work) into two clouds for the Paper III analysis, and excluding leaves that are isolated and not part of the CMZ hierarchy. The final molecular cloud catalog is also re-numbered to be ordered with Galactic longitude. The molecular cloud catalog is described in detail in Paper III. We include in Table \ref{tab:general_properties} the relevant Paper III cloud IDs for easy inter-comparison. We also list the approximate region IDs from the CMZoom Survey \citep{Battersby2020, Hatchfield2020} for comparison.

The derivation of these physical properties is described in Section \ref{sec:physical_properties}, the kinematic properties in Section \ref{sec:kinematic_properties}, and the luminosities and SFRs in Section \ref{sec:luminosity}.
All three tables, Tables \ref{tab:general_properties}, \ref{tab:kinematic_properties}, \ref{tab:luminosity_properties} have identical row organization. The structures are separated by whether they are categorized as branches (listed first) or leaves (listed second). Then, within each category, the structures are listed in order of decreasing total structure area.

We release all the analysis products from this series of papers (including the dendrogram masks and a full machine-readable table including all columns from Tables \ref{tab:general_properties}-\ref{tab:powerlaw})
in the 3-D CMZ Harvard Dataverse: \href{https://dataverse.harvard.edu/dataverse/3D_CMZ}{https://dataverse.harvard.edu/dataverse/3D\textunderscore CMZ}. The dataset for this paper is found in the Papers I and II repository: \href{https://doi.org/10.7910/DVN/7DOJG5}{https://doi.org/10.7910/DVN/7DOJG5}. The Harvard Dataverse is an online data repository that enables long-term data preservation and sharing. Project updates can also be found on the 3-D CMZ website:  \href{https://centralmolecularzone.github.io/3D_CMZ/}{https://centralmolecularzone.github.io/3D\textunderscore CMZ/}.

\subsection{Physical Properties of CMZ Structures}
\label{sec:physical_properties}

Most of the calculations of physical properties tabulated in Table \ref{tab:general_properties} are fairly straightforward, but we describe here some of the details and nuances. 
The source name is from the source coordinates which is the dendrogram-specified purely geometric mean position of the overall structure mask in x and y. The structure ID is computed by the dendrogram algorithm and does not have physical meaning. We also include a colloquiual name or description of key structures in the final column of Table \ref{tab:general_properties}. For detailed exploration of the catalog or comparison with other data, please download the dendrogram mask directly from our Dataverse (see link at the end of Section \ref{sec:tables}).

The area for each structure is simply the number of pixels contained in the dendrogram mask multiplied by the pixel area at a Galactic Center distance of 8.2 kpc (assumed throughout this work). The median and peak column densities for each structure are from the Herschel column density map from Paper I. The mass is the average column density of each structure multiplied by its area. As described in Paper I, this mass is therefore our best estimate of the total molecular gas. 

We report a radius for each structure, which is defined as the radius of a circle whose area equals the total dendrogram mask area. The volume density is calculated assuming that this radius (which is derived from a 2-D plane-of-sky projection) corresponds to the true average 3-D radius of the structure. We divide the total structure mass by the volume of a 3-D sphere with this radius. This assumption is reasonable for leaves, but not very realistic for the branches and other large structures in the hierarchy. We report both a median and peak dust temperature within each structure from the dust temperature map in Paper I.

\subsection{Kinematic Properties of CMZ Structures}
\label{sec:kinematic_properties}

We report simple kinematic properties of each dendrogram structure using 3 mm data from the Mopra radio telescope from \citet{Jones2012} in Table \ref{tab:kinematic_properties}. We report the propreties of three spectral lines: HNCO 4(0,4) - 3(0,3) at 87.925238 GHz, HCN 1-0 (F=2-1) at 88.6318473 GHz, and HC$_3$N 10-9 at 90.978989 GHz. These molecular line transitions tell us different things about the molecular gas. This combination was  chosen to represent both the variation in properties one expects, as well as for comparison with lines observed in other regions. 

We first extract an integrated spectra for each line over each dendrogram structure mask. We compute the moment 0 (integrated value), moment 1 (the intensity weighted velocity), and moment 2 (intensity weighted velocity dispersion) of the integrated spectra for each line. In Table \ref{tab:kinematic_properties} we report these values. In the case of the moment 2 values, we have converted from the native velocity dispersion $\sigma$ to the equivalent FWHM (full width at half maximum) for a Gaussian line profile.

The CMZ is a complex region, both kinematically and chemically, with emission from different regions along many lines-of-sight.
For example, Figure \ref{fig:dendro} shows the HC$_3$N spectra for structures 46 and 36 (the Dust Ridge Bridge and the 20 \kms~cloud). While structure 36 (20\kms~cloud) shows a single wide, somewhat asymmetric spectral line centered around 18\kms, structure 46 (Dust Ridge Bridge) shows one strong peak centered at about 55\kms~which dominates the moment calculation and a second small peak around -30\kms. This likely indicates that structure 46 contains gas from more than one line-of-sight location in the CMZ while structure 36 is a single cloud structure that is well-suited to moment analysis. This complexity is well-documented in e.g. \citet{Henshaw2016b}, \citet{Shetty2012} or \citet{ Kauffmann2017a}, who perform more comprehensive kinematic analyses of gas in the CMZ.

Our approach of using moment analysis is chosen to enable study of the entire CMZ hierarchy in a consistent and simple way. It enables comparison with other integrated structure moment analysis works from Milky Way and extragalactic observations, which we show in Section \ref{sec:exgal}.  
The third paper in this series, Paper III, carefully and systematically refines this catalog into well-defined molecular clouds in position-position-velocity space. Paper III performs a detailed kinematic analysis of each of these clouds, including multiple component Gaussian fitting using pyspeckit, which implements the Levenberg-Marquardt algorithm.

\begin{table*}
\centering
\caption{Kinematic properties of the dendrogram structures. Shown for each structure is the source name, leaf ID, and moment-analysis-computed integrated intensity (mom0), centroid velocity (mom1), and velocity full-width at half maximum (FWHM), for HNCO 4$_{0,4}$-3$_{0,3}$, HCN 1-0, and HC$_{3}$N 10-9. The grouping and order of structures is the same as in Table \ref{tab:general_properties}.}\resizebox{\linewidth}{!}{%
\begin{tabular}{ccccccccccc}
\hline
Name & ID & \multicolumn{3}{c}{Integrated intensity ($\mathrm{K\,km\,s^{-1}}$)} & \multicolumn{3}{c}{Weighted Velocity ($\mathrm{km\,s^{-1}}$)} & \multicolumn{3}{c}{FWHM ($\mathrm{km\,s^{-1}}$)} \\ \hline
     &  & HNCO 4$_{0,4}$-3$_{0,3}$ & HCN 1-0 & HC$_{3}$N 10-9 & HNCO 4$_{0,4}$-3$_{0,3}$ & HCN 1-0 & HC$_{3}$N 10-9 & HNCO 4$_{0,4}$-3$_{0,3}$ & HCN 1-0 & HC$_{3}$N 10-9 \\ \hline
    Branches & & & & & & & & &\\ \hline
    G0.720-0.069 & 8 & 1.4E-02 & 6.3E-02 & 6.0E-03 & 57 & 67 & 51 & 72 & 95 & 79 \\
    G0.750-0.070 & 9 & 1.6E-02 & 6.8E-02 & 6.6E-03 & 56 & 66 & 50 & 70 & 94 & 76 \\
    G0.970-0.067 & 11 & 2.1E-02 & 7.2E-02 & 7.7E-03 & 58 & 72 & 54 & 67 & 88 & 73 \\
    G0.840-0.065 & 12 & 2.2E-02 & 8.1E+01 & 9.2E+00 & 59 & 73 & 53 & 73 & 88 & 75 \\
    G0.830-0.062 & 13 & 2.4E-02 & 8.3E+01 & 9.7E+00 & 58 & 72 & 53 & 73 & 88 & 74 \\
    G0.820-0.059 & 14 & 2.9E-02 & 8.8E+01 & 1.1E+01 & 57 & 71 & 52 & 71 & 87 & 72 \\
    G0.800-0.060 & 15 & 3.3E-02 & 9.2E+01 & 1.3E+01 & 55 & 70 & 51 & 68 & 87 & 68 \\
    G359.900-0.071 & 19 & 8.9E-03 & 7.6E+01 & 6.9E+00 & 32 & 45 & 36 & 84 & 94 & 74 \\
    G0.020-0.058 & 27 & 1.2E-02 & 9.5E+01 & 9.0E+00 & 34 & 48 & 37 & 72 & 83 & 67 \\
    G1.610-0.059 & 23 & 1.6E-02 & 4.4E-02 & 4.1E-03 & 51 & 64 & 53 & 26 & 67 & 30 \\
    G0.670-0.059 & 17 & 4.2E-02 & 1.1E+02 & 1.8E+01 & 46 & 63 & 46 & 49 & 87 & 52 \\
    G0.040-0.054 & 31 & 1.5E-02 & 1.1E+02 & 1.1E+01 & 34 & 47 & 37 & 65 & 77 & 61 \\
    G0.650-0.040 & 25 & 4.7E-02 & 1.1E+02 & 2.1E+01 & 49 & 63 & 48 & 48 & 86 & 52 \\
    G1.040-0.051 & 29 & 2.7E-02 & 8.1E+01 & 7.5E+00 & 83 & 80 & 81 & 24 & 72 & 32 \\
    G359.530-0.120 & 18 & 7.2E-03 & 5.1E+01 & 4.9E+00 & -55 & -45 & -50 & 37 & 150 & 80 \\
    G0.670-0.038 & 26 & 5.6E-02 & 1.2E+02 & 2.5E+01 & 53 & 66 & 53 & 43 & 84 & 45 \\
    G359.960-0.078 & 32 & 2.0E-02 & 1.2E+02 & 1.6E+01 & 29 & 43 & 34 & 61 & 72 & 58 \\
    G0.660-0.023 & 37 & 6.5E-02 & 1.2E+02 & 3.0E+01 & 58 & 73 & 59 & 31 & 80 & 29 \\
    G0.210-0.003 & 44 & 1.3E-02 & 9.9E+01 & 7.9E+00 & 46 & 54 & 44 & 64 & 78 & 63 \\
    G359.910-0.079 & 34 & 2.6E-02 & 1.3E+02 & 2.1E+01 & 22 & 37 & 28 & 48 & 73 & 56 \\
    G0.220-0.490 & 3 & - & - & - & - & - & - & - & - & - \\
    G0.110-0.079 & 35 & 2.1E-02 & 1.4E+02 & 1.3E+01 & 53 & 52 & 52 & 25 & 56 & 23 \\
    G0.400+0.048 & 50 & 1.8E-02 & 6.0E+01 & 8.6E+00 & 17 & 28 & 16 & 32 & 43 & 38 \\
    \hline
    Leaves & & & & & & & & &\\ \hline
    G359.510-0.130 & 24 & 9.7E-03 & 5.6E+01 & 5.6E+00 & -56 & -42 & -55 & 22 & 130 & 26 \\
    G1.070-0.049 & 33 & 2.7E-02 & 6.9E+01 & 6.1E+00 & 84 & 79 & 79 & 21 & 71 & 32 \\
    G0.230-0.004 & 45 & 1.4E-02 & 9.3E+01 & 7.1E+00 & 42 & 53 & 35 & 70 & 81 & 65 \\
    G0.890-0.044 & 39 & 3.5E-02 & 9.0E+01 & 9.7E+00 & 83 & 77 & 48 & 20 & 68 & 120 \\
    G0.820-0.190 & 21 & 4.3E-02 & 1.0E+02 & 1.6E+01 & 39 & 69 & 40 & 34 & 86 & 33 \\
    G1.720-0.390 & 6 & - & - & - & - & - & - & - & - & - \\
    G0.670-0.028 & 43 & 8.1E-02 & 1.3E+02 & 4.2E+01 & 62 & 76 & 61 & 29 & 76 & 27 \\
    G0.720-0.090 & 30 & 6.0E-02 & 1.3E+02 & 2.6E+01 & 39 & 59 & 32 & 54 & 82 & 50 \\
    G359.690-0.130 & 28 & 6.4E-03 & 4.0E+01 & 2.4E+00 & -26 & 4 & -27 & 24 & 140 & 16 \\
    G359.880-0.081 & 36 & 3.8E-02 & 1.2E+02 & 2.6E+01 & 15 & 24 & 15 & 27 & 55 & 29 \\
    G1.020-0.230 & 16 & 5.9E-03 & 4.0E+01 & 1.5E+00 & 97 & 96 & 92 & 55 & 75 & 54 \\
    G0.550-0.870 & 0 & - & - & - & - & - & - & - & - & - \\
    G359.600-0.220 & 20 & 5.4E-03 & 5.8E+01 & 7.1E+00 & -13 & -20 & -34 & 95 & 170 & 120 \\
    G0.240-0.470 & 5 & - & - & - & - & - & - & - & - & - \\
    G0.490+0.008 & 48 & 3.3E-02 & 9.0E+01 & 1.3E+01 & 28 & 40 & 30 & 25 & 55 & 24 \\
    G0.120+0.003 & 46 & 1.3E-02 & 1.0E+02 & 1.2E+01 & 52 & 56 & 54 & 26 & 64 & 28 \\
    G1.600+0.012 & 47 & 3.1E-02 & 5.3E+01 & 5.8E+00 & 53 & 89 & 52 & 23 & 150 & 19 \\
    G1.650-0.052 & 42 & 2.9E-02 & 3.9E+01 & 7.5E+00 & 50 & 59 & 51 & 15 & 23 & 12 \\
    G359.940+0.160 & 54 & 6.8E-04 & 1.0E+01 & 5.3E-01 & -110 & 55 & 0 & 100 & 140 & 4 \\
    G0.330-0.190 & 22 & 1.4E-03 & 1.6E+01 & 5.7E-01 & 16 & 20 & 18 & 29 & 12 & 5 \\
    G3.430-0.350 & 10 & - & - & - & - & - & - & - & - & - \\
    G0.200-0.520 & 4 & - & - & - & - & - & - & - & - & - \\
    G0.410+0.048 & 51 & 2.4E-02 & 6.6E+01 & 1.0E+01 & 19 & 28 & 18 & 25 & 38 & 28 \\
    G359.980-0.071 & 41 & 3.3E-02 & 1.8E+02 & 3.8E+01 & 48 & 53 & 47 & 25 & 58 & 26 \\
    G0.340+0.060 & 53 & 7.9E-03 & 3.8E+01 & 5.3E+00 & -2 & 19 & -3 & 34 & 60 & 38 \\
    G0.120-0.081 & 38 & 2.6E-02 & 1.5E+02 & 1.6E+01 & 54 & 54 & 53 & 22 & 54 & 22 \\
    G358.460-0.390 & 7 & - & - & - & - & - & - & - & - & - \\
    G356.660+0.560 & 56 & - & - & - & - & - & - & - & - & - \\
    G0.070-0.076 & 40 & 3.0E-02 & 1.7E+02 & 2.2E+01 & 50 & 51 & 51 & 22 & 50 & 21 \\
    G0.090-0.660 & 2 & - & - & - & - & - & - & - & - & - \\
    G0.380+0.050 & 52 & 1.1E-02 & 5.1E+01 & 5.5E+00 & 14 & 28 & 11 & 49 & 47 & 52 \\
    G356.520+0.210 & 55 & - & - & - & - & - & - & - & - & - \\
    G357.070-0.770 & 1 & - & - & - & - & - & - & - & - & - \\
    G0.650+0.030 & 49 & 7.4E-02 & 1.3E+02 & 2.7E+01 & 53 & 67 & 52 & 28 & 70 & 30 \\
\hline \hline
\end{tabular}%
\label{tab:kinematic_properties}
}
\end{table*}

 \begin{table*}
\centering
\caption{Luminosity, SFR, and confinement properties of the dendrogram structures. Shown for each structure is the source name, structure ID, total luminosity and mean luminosity for the cool and warm components, respectively, the combined total IR luminosity, the SFR inferred from that total IR luminosity assuming the star-forming relation from \citet{Kennicutt1998}, the free-fall time, the SFR estimated using the method from \citet{Barnes2017} but using the free-fall time, the SFR reported in matching sources from the CMZoom catalog \citep{Hatchfield2024}, the best overall SFR estimate for each structure, the external pressure requirement for confinement, the total pressure of the parent structure in the dendrogram, the ratio of the total parent pressure / required external pressure, and whether a leaf is confined by the pressure of it's parent structure (Yes/No/Self-gravitating). The grouping and order of structures is the same as in Table \ref{tab:general_properties}. $^{*}$See section \ref{sec:sfr} for details regarding the choice of best SFR estimate. No best SFR estimate is given for branches with A \textless\ 1000~pc$^{-2}$.}\resizebox{\linewidth}{!}{%
\begin{tabular}{ccccccccccccccccc}
\hline
Name & ID & $L_{\textrm{tot}}$ & $\langle L \rangle$ & $L_{\textrm{tot}}$ & $\langle L \rangle$ & $L_{\textrm{total IR}}$ & $\textrm{SFR}_{\textrm{total IR}}$ & $t_{\textrm{ff}}$ & $\textrm{SFR}_{\textrm{FF}}$ & $\textrm{SFR}_{\textrm{CMZoom}}$ & $\textrm{SFR}_{\textrm{CMZoom}}$ & \textbf{Best SFR} & Required & Parent & Pressure & Confined\\
& & (Cool) & (Cool) & (Warm) & (Warm) & & & & & Herschel temp & 50K  &\textbf{Estimate$^{*}$} & Ext P/$k$ & Tot P/$k$ & Ratio & \\ \hline
& & $\mathrm{L_{\odot}}$ & $\mathrm{L_{\odot}}$/pix & $\mathrm{L_{\odot}}$ & $\mathrm{L_{\odot}}$/pix & $\mathrm{L_{\odot}}$ & $\mathrm{M_{\odot}\,yr^{-1}}$ & Myr & $\mathrm{M_{\odot}\,yr^{-1}}$ & $\mathrm{M_{\odot}\,yr^{-1}}$ & $\mathrm{M_{\odot}\,yr^{-1}}$ & $\mathrm{M_{\odot}\,yr^{-1}}$ & K cm$^{-3}$ & K cm$^{-3}$ & & \\ \hline
Branches & \\ \hline
G0.720-0.069 & 8 & 1.40E+08 & 2.20E+03 & 6.10E+07 & 1.70E+03 & 2.10E+08 & 3.50E-02 & - & - & - & - & 3.50E-02 & - & - & - & - \\
G0.750-0.070 & 9 & 1.20E+08 & 2.30E+03 & 4.90E+07 & 1.70E+03 & 1.70E+08 & 3.00E-02 & - & - & - & - & 3.00E-02 & - & - & - & - \\
G0.970-0.067 & 11 & 5.90E+07 & 1.70E+03 & 1.40E+07 & 8.80E+02 & 7.30E+07 & 1.30E-02 & - & - & - & - & 1.30E-02 & - & - & - & - \\
G0.840-0.065 & 12 & 5.10E+07 & 2.10E+03 & 1.20E+07 & 8.60E+02 & 6.40E+07 & 1.10E-02 & - & - & - & - & 1.10E-02 & - & - & - & - \\
G0.830-0.062 & 13 & 4.90E+07 & 2.20E+03 & 1.20E+07 & 8.60E+02 & 6.10E+07 & 1.00E-02 & - & - & - & - & 1.00E-02 & - & - & - & - \\
G0.820-0.059 & 14 & 4.00E+07 & 2.30E+03 & 8.50E+06 & 7.90E+02 & 4.90E+07 & 8.40E-03 & - & - & - & - & 8.40E-03 & - & - & - & - \\
G0.800-0.060 & 15 & 3.30E+07 & 2.60E+03 & 6.50E+06 & 7.70E+02 & 4.00E+07 & 6.80E-03 & - & - & - & - & 6.80E-03 & - & - & - & - \\
G359.900-0.071 & 19 & 4.70E+07 & 4.70E+03 & 2.50E+07 & 3.00E+03 & 7.20E+07 & 1.20E-02 & - & - & - & - & 1.20E-02 & - & - & - & - \\
G0.020-0.058 & 27 & 3.60E+07 & 5.20E+03 & 1.90E+07 & 3.60E+03 & 5.40E+07 & 9.30E-03 & - & - & - & - & 9.30E-03 & - & - & - & - \\
G1.610-0.059 & 23 & 3.60E+06 & 5.50E+02 & - & - & 3.60E+06 & 6.20E-04 & - & - & - & - & 6.20E-04 & - & - & - & - \\
G0.670-0.059 & 17 & 2.40E+07 & 3.90E+03 & 4.90E+06 & 8.10E+02 & 2.90E+07 & 4.90E-03 & - & - & - & - & 4.90E-03 & - & - & - & - \\
G0.040-0.054 & 31 & 2.70E+07 & 5.80E+03 & 1.30E+07 & 4.10E+03 & 4.00E+07 & 6.90E-03 & - & - & - & - & - & - & - & - & - \\
G0.650-0.040 & 25 & 1.90E+07 & 4.70E+03 & 3.40E+06 & 8.80E+02 & 2.20E+07 & 3.80E-03 & - & - & - & - & - & - & - & - & - \\
G1.040-0.051 & 29 & 4.80E+06 & 1.40E+03 & 5.90E+05 & 4.80E+02 & 5.40E+06 & 9.40E-04 & - & - & - & - & - & - & - & - & - \\
G359.530-0.120 & 18 & 9.30E+06 & 3.60E+03 & 4.90E+06 & 1.90E+03 & 1.40E+07 & 2.40E-03 & - & - & - & - & - & - & - & - & - \\
G0.670-0.038 & 26 & 1.30E+07 & 5.40E+03 & 2.00E+06 & 8.50E+02 & 1.50E+07 & 2.50E-03 & - & - & - & - & - & - & - & - & - \\
G359.960-0.078 & 32 & 1.30E+07 & 5.80E+03 & 6.90E+06 & 3.90E+03 & 2.00E+07 & 3.40E-03 & - & - & - & - & - & - & - & - & - \\
G0.660-0.023 & 37 & 7.90E+06 & 7.10E+03 & 1.10E+06 & 9.80E+02 & 8.90E+06 & 1.50E-03 & - & - & - & - & - & - & - & - & - \\
G0.210-0.003 & 44 & 5.60E+06 & 5.10E+03 & 2.90E+06 & 4.20E+03 & 8.50E+06 & 1.50E-03 & - & - & - & - & - & - & - & - & - \\
G359.910-0.079 & 34 & 6.80E+06 & 6.30E+03 & 3.10E+06 & 3.60E+03 & 9.90E+06 & 1.70E-03 & - & - & - & - & - & - & - & - & - \\
G0.220-0.490 & 3 & 2.10E+05 & 3.70E+02 & 2.80E+05 & 6.60E+02 & 4.90E+05 & 8.50E-05 & - & - & - & - & - & - & - & - & - \\
G0.110-0.079 & 35 & 1.70E+06 & 4.00E+03 & 1.60E+06 & 4.40E+03 & 3.30E+06 & 5.70E-04 & - & - & - & - & - & - & - & - & - \\
G0.400+0.048 & 50 & 8.50E+05 & 3.30E+03 & 1.90E+05 & 7.60E+02 & 1.00E+06 & 1.80E-04 & - & - & - & - & - & - & - & - & - \\ \hline
Leaves & \\ \hline
G359.510-0.130 & 24 & 4.50E+06 & 3.90E+03 & 2.20E+06 & 1.90E+03 & 6.70E+06 & 1.20E-03 & 7.89E-01 & 7.70E-03 & - & - & 7.70E-03 & 2.80E+07 & 8.00E+07 & 2.88 & Y \\
G1.070-0.049 & 33 & 1.50E+06 & 1.40E+03 & 3.80E+04 & 4.20E+02 & 1.60E+06 & 2.70E-04 & 5.94E-01 & 4.90E-03 & 7.10E-03 & 1.10E-03 & 7.10E-03 & 2.20E+07 & 1.80E+08 & 8.23 & Y \\
G0.230-0.004 & 45 & 4.10E+06 & 5.10E+03 & 2.70E+06 & 4.20E+03 & 6.80E+06 & 1.20E-03 & 6.40E-01 & 9.50E-03 & - & - & 9.50E-03 & 6.80E+08 & 2.70E+08 & 0.40 & N \\
G0.890-0.044 & 39 & 1.30E+06 & 1.80E+03 & 3.40E+05 & 4.70E+02 & 1.60E+06 & 2.80E-04 & 5.39E-01 & 5.40E-03 & - & - & 5.40E-03 & 3.00E+07 & 1.80E+08 & 6.10 & Y \\
G0.820-0.190 & 21 & 1.30E+06 & 2.10E+03 & 2.80E+05 & 4.50E+02 & 1.60E+06 & 2.80E-04 & 5.26E-01 & 5.60E-03 & - & - & 5.60E-03 & 1.90E+08 & 5.10E+08 & 2.63 & Y \\
G1.720-0.390 & 6 & 1.50E+05 & 2.70E+02 & - & - & 1.50E+05 & 2.60E-05 & 7.20E-01 & 1.00E-03 & - & - & 1.00E-03 & - & - & - & - \\
G0.670-0.028 & 43 & 4.90E+06 & 9.60E+03 & 4.70E+05 & 9.50E+02 & 5.40E+06 & 9.20E-04 & 2.11E-01 & 2.60E-02 & - & - & 2.60E-02 & -5.10E+08 & 2.40E+09 & - & SG \\
G0.720-0.090 & 30 & 1.90E+06 & 3.90E+03 & 3.00E+05 & 6.00E+02 & 2.20E+06 & 3.90E-04 & 3.98E-01 & 8.70E-03 & 1.40E-02 & 2.70E-03 & 1.40E-02 & 9.70E+08 & 1.50E+09 & 1.51 & Y \\
G359.690-0.130 & 28 & 1.40E+06 & 2.90E+03 & 7.70E+05 & 1.60E+03 & 2.20E+06 & 3.70E-04 & 7.44E-01 & 4.70E-03 & - & - & 4.70E-03 & 5.10E+07 & 1.40E+08 & 2.76 & Y \\
G359.880-0.081 & 36 & 1.90E+06 & 5.10E+03 & 8.50E+05 & 2.40E+03 & 2.80E+06 & 4.80E-04 & 3.33E-01 & 1.20E-02 & 5.10E-02 & 9.20E-03 & 9.20E-03 & 2.00E+08 & 4.60E+08 & 2.26 & Y \\
G1.020-0.230 & 16 & 4.20E+05 & 1.10E+03 & 8.40E+04 & 5.30E+02 & 5.00E+05 & 8.60E-05 & 7.05E-01 & 2.20E-03 & - & - & 2.20E-03 & 3.40E+08 & 1.60E+08 & 0.47 & N \\
G0.550-0.870 & 0 & 6.10E+05 & 1.80E+03 & - & - & 6.10E+05 & 1.10E-04 & 6.33E-01 & 2.70E-03 & - & - & 2.70E-03 & - & - & - & - \\
G359.600-0.220 & 20 & 8.60E+05 & 2.70E+03 & 4.80E+05 & 1.50E+03 & 1.30E+06 & 2.30E-04 & 5.74E-01 & 4.60E-03 & - & - & 4.60E-03 & 1.60E+09 & 8.00E+07 & 0.05 & N \\
G0.240-0.470 & 5 & 1.10E+05 & 4.00E+02 & 1.90E+05 & 6.70E+02 & 3.00E+05 & 5.20E-05 & 7.04E-01 & 1.60E-03 & - & - & 1.60E-03 & - & - & - & - \\
G0.490+0.008 & 48 & 1.00E+06 & 4.00E+03 & 1.80E+05 & 7.10E+02 & 1.20E+06 & 2.10E-04 & 3.07E-01 & 8.20E-03 & 1.70E-02 & 2.90E-03 & 1.70E-02 & 2.20E+08 & 8.10E+08 & 3.75 & Y \\
G0.120+0.003 & 46 & 1.20E+06 & 5.00E+03 & - & - & 1.20E+06 & 2.10E-04 & 5.32E-01 & 4.70E-03 & 1.70E-03 & 5.00E-04 & 1.70E-03 & 1.20E+08 & 2.70E+08 & 2.23 & Y \\
G1.600+0.012 & 47 & 1.50E+05 & 6.80E+02 & - & - & 1.50E+05 & 2.60E-05 & 3.73E-01 & 2.00E-03 & 5.00E-03 & 5.00E-04 & 5.00E-03 & 1.60E+08 & 6.60E+07 & 0.42 & N \\
G1.650-0.052 & 42 & 1.10E+05 & 5.20E+02 & - & - & 1.10E+05 & 1.90E-05 & 3.73E-01 & 1.60E-03 & 1.50E-03 & 2.00E-04 & 1.50E-03 & 3.20E+07 & 6.60E+07 & 2.06 & Y \\
G359.940+0.160 & 54 & 1.70E+05 & 9.20E+02 & 1.60E+05 & 8.20E+02 & 3.30E+05 & 5.70E-05 & 6.06E-01 & 1.90E-03 & - & - & 1.90E-03 & 1.60E+09 & 8.30E+07 & 0.05 & N \\
G0.330-0.190 & 22 & 2.90E+05 & 1.80E+03 & 3.20E+05 & 2.20E+03 & 6.00E+05 & 1.00E-04 & 6.28E-01 & 2.70E-03 & - & - & 2.70E-03 & - & - & - & - \\
G3.430-0.350 & 10 & 5.50E+04 & 3.30E+02 & - & - & 5.50E+04 & 9.50E-06 & 5.83E-01 & 6.90E-04 & - & - & 6.90E-04 & - & - & - & - \\
G0.200-0.520 & 4 & 5.40E+04 & 3.40E+02 & 1.50E+04 & 4.70E+02 & 6.90E+04 & 1.20E-05 & 5.77E-01 & 7.90E-04 & - & - & 7.90E-04 & - & - & - & - \\
G0.410+0.048 & 51 & 4.00E+05 & 3.20E+03 & 7.90E+04 & 6.50E+02 & 4.80E+05 & 8.20E-05 & 3.28E-01 & 4.50E-03 & 2.70E-03 & 5.00E-04 & 2.70E-03 & 2.80E+08 & 2.80E+08 & 1.03 & Y \\
G359.980-0.071 & 41 & 9.10E+05 & 8.30E+03 & 4.10E+05 & 5.50E+03 & 1.30E+06 & 2.30E-04 & 2.79E-01 & 9.40E-03 & 6.30E-03 & 1.80E-03 & 6.30E-03 & 3.60E+08 & 4.60E+08 & 1.28 & Y \\
G0.340+0.060 & 53 & 3.50E+05 & 3.50E+03 & 1.20E+05 & 1.20E+03 & 4.70E+05 & 8.10E-05 & 4.14E-01 & 3.50E-03 & - & - & 3.50E-03 & 3.70E+08 & 1.90E+08 & 0.51 & N \\
G0.120-0.081 & 38 & 4.30E+05 & 4.30E+03 & 3.60E+05 & 3.60E+03 & 7.90E+05 & 1.40E-04 & 3.12E-01 & 6.30E-03 & 4.00E-04 & 1.00E-04 & 4.00E-04 & 2.30E+08 & 1.90E+08 & 0.83 & N \\
G358.460-0.390 & 7 & 3.80E+04 & 4.70E+02 & - & - & 3.80E+04 & 6.60E-06 & 4.72E-01 & 7.00E-04 & - & - & 7.00E-04 & - & - & - & - \\
G356.660+0.560 & 56 & 1.80E+03 & 3.00E+01 & - & - & 1.80E+03 & 3.20E-07 & 5.28E-01 & 2.00E-04 & - & - & 2.00E-04 & - & - & - & - \\
G0.070-0.076 & 40 & 2.20E+05 & 3.80E+03 & 2.90E+05 & 6.40E+03 & 5.10E+05 & 8.80E-05 & 2.50E-01 & 6.10E-03 & 6.40E-03 & 1.50E-03 & 6.40E-03 & 3.50E+08 & 1.90E+08 & 0.54 & N \\
G0.090-0.660 & 2 & 2.80E+04 & 5.30E+02 & - & - & 2.80E+04 & 4.80E-06 & 4.65E-01 & 6.10E-04 & - & - & 6.10E-04 & - & - & - & - \\
G0.380+0.050 & 52 & 1.70E+05 & 3.70E+03 & 3.70E+04 & 8.00E+02 & 2.10E+05 & 3.60E-05 & 2.74E-01 & 3.30E-03 & 3.90E-02 & 1.00E-02 & 1.00E-02 & 1.80E+09 & 2.80E+08 & 0.15 & N \\
G356.520+0.210 & 55 & 1.80E+03 & 4.70E+01 & - & - & 1.80E+03 & 3.20E-07 & 4.65E-01 & 2.30E-04 & - & - & 2.30E-04 & - & - & - & - \\
G357.070-0.770 & 1 & 3.50E+02 & 1.30E+01 & - & - & 3.50E+02 & 6.00E-08 & 4.30E-01 & - & - & - & 6.00E-08 & - & - & - & - \\
G0.650+0.030 & 49 & 1.10E+05 & 3.80E+03 & 1.20E+04 & 3.90E+02 & 1.30E+05 & 2.20E-05 & 1.64E-01 & 4.10E-03 & - & - & 4.10E-03 & 1.50E+09 & 2.40E+09 & 1.65 & Y \\ \hline \hline
\end{tabular}%
\label{tab:luminosity_properties}
}
\end{table*}

\subsection{Luminosities of CMZ Structures}
\label{sec:luminosity}

In Paper I of this series, we computed modified blackbody fits at each source pixel within the inner $\ell$ = 40\deg~of the Galaxy. In fitting a single-component modified blackbody function to each position, we assumed that the dust emission was well-fit by a single temperature component. The wavelengths included in the fits were 160, 250, 350, and 500 \micron, thereby making this measurement most sensitive to relatively ``cool" dust ($\sim 20$K). We integrate underneath the modified blackbody fit at each source pixel to generate a ``cool" far-IR luminosity map of the region. We then integrate the luminosity values in this map over each of our Herschel dendrogram structure masks and report an integrated as well as structure average ``cool" luminosity in Table \ref{tab:luminosity_properties}.

In one of our earlier papers, \citet{Barnes2017}, we performed two temperature component modified blackbody fits in the inner 2\deg~longitude and 1\deg~latitude of the Galaxy. In these fits, we also incorporated shorter wavelength data from Spitzer, making these measurements more sensitive to an additional ``warm" component of the dust. Since \citep{Barnes2017} used the same background-subtraction method as this work, we use the ``warm" IR dust component map from \citet{Barnes2017} to calculate and report a total and average ``warm" IR luminosity for each dendrogram structure in Table \ref{tab:luminosity_properties}.
We note that since the two-component ``warm" temperature component analysis was only performed in the inner 2\deg~longitude and 1\deg~latitude of the Galaxy, structures outside these boundaries (including if only part of the structure is outside the boundary) are excluded in the ``warm" luminosity column in the Table. 

In Table \ref{tab:luminosity_properties} we report both the ``cool" and ``warm" IR luminosities  for each structure. Additionally, we add together these two luminosity components for each structure and report the total IR luminosity.

% Made in python - SFR_comparisons_megatable.ipynb
\begin{figure*}
\centering
\includegraphics[width=1\textwidth]{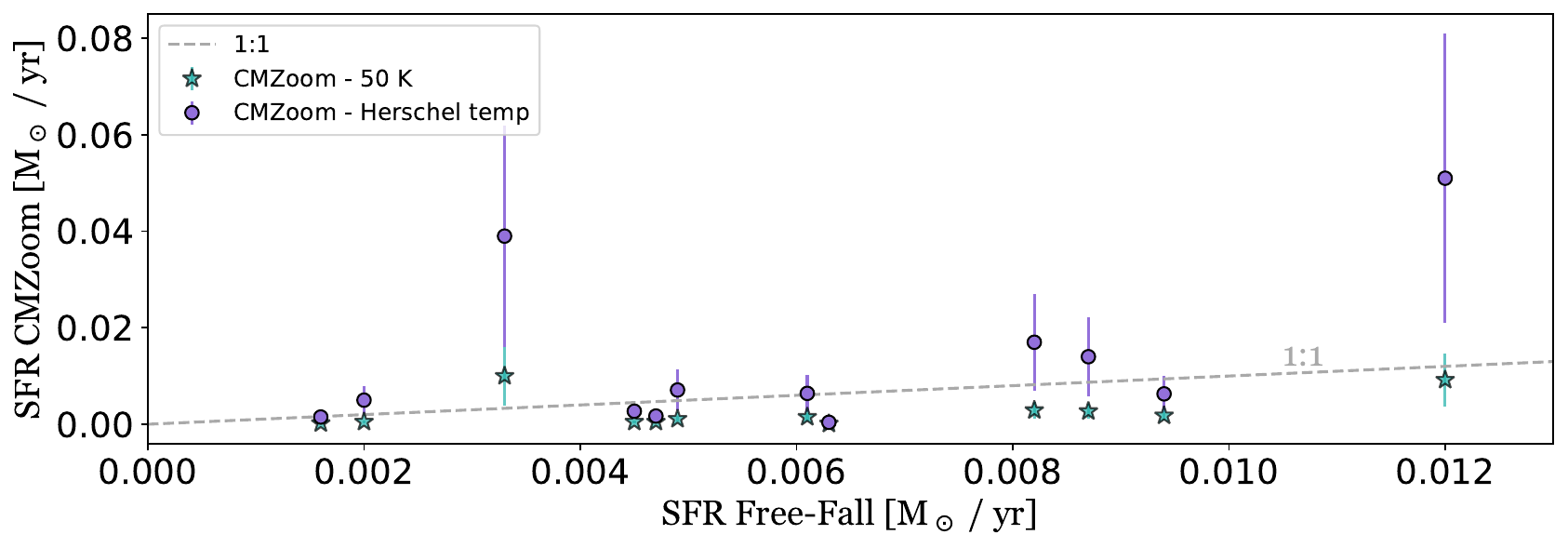}%SFR_YLOG_cmzoom_comparison_v2.pdf
\caption{A comparison between the SFRs calculated from CMZoom \citep[][shown on the y-axis]{Hatchfield2024} and using the free-fall method \citep[this work in Section \ref{sec:sfr}, based on][shown on the x-axis]{Barnes2017} shows generally good agreement in the 12 sources for which both measurements are available. The dashed gray line shows the 1:1 line. The purple circles show the SFR estimates from CMZoom assuming the Herschel-calculated dust temperature (our default SFR choice), while the green stars from CMZoom assume a higher dust temperature of 50 K \citep{Hatchfield2024}. Including their 1-$\sigma$ error bars, most of the points agree with the 1:1 line, with two clear outliers that have much higher CMZoom estimated SFRs than their free-fall estimates. From left to right in this plot, the two high outliers are cloud c (G0.380$+$0.050) and the 20\kms~cloud (G359.880-0.081). These two regions are actively star-forming, so the 50 K SFR estimate from CMZoom is likely more appropriate (see Section \ref{sec:sfr}) and is therefore used in Table \ref{tab:luminosity_properties} and agrees better with the free-fall SFR estimate. The very low outlier is the Straw and Sticks clouds (ID: 38). Table \ref{tab:luminosity_properties} reports all calculated SFRs for each source as well as indicating in the final column the best available SFR, which is discussed in more detail in Section \ref{sec:sfr}.}
\label{fig:sfr_comparison}
\end{figure*}

\subsection{Estimates of the Star Formation Rates of CMZ Structures}
\label{sec:sfr}

We estimate SFRs in each CMZ structure. We implement three different methods to do this which are most relevant on different size scales. The first is using the total IR luminosity and can be applied to only the largest structures in our sample. The second two are intended to be used on individual molecular clouds in the CMZ. Each method is described in more detail below.

Building upon the work from \citet{Barnes2017}, we use the total IR luminosity calculated in Section \ref{sec:luminosity} to estimate a total star formation rate (SFR) for each structure by implementing the conversion factor from \citet{Kennicutt1998} of 4.5 $\times 10^{-44}$ L (TIR) erg s$^{-1}$ \Msun$^{-1}$ yr. This value is reported in Table \ref{tab:luminosity_properties}.

While using the total IR luminosity is a common method to estimate global SFRs on large scales, this calibration assumes a large and complete stellar population with regions in a range of evolutionary states and is not valid for the smaller structures in our sample. Additionally, it assumes that the total IR for each structures is produced by internal heating, which is not the case for some CMZ molecular clouds, such as the ``Brick" which is predominantly heated by the external radiation field. Therefore, we implement two additional methods to estimate SFRs for the small structures in our sample. These methods are described below. In our analysis in this paper, we only utilize the global SFR estimates from total IR luminosity for branches in our catalog with a total structure surface area of 1000 pc$^2$ or greater. This cutoff is driven by the idea that on scales of $>$ 1 kpc, one can generally assume a complete stellar population in a variety of evolutionary stages, so this SFR estimate should be valid for structures larger than 1000 pc$^2$ \citep{Kennicutt1998}. \citet{Barnes2017} compare this total IR luminosity SFR in the entire CMZ with other measures of the SFR based on IR luminosity, YSO counting, and free-free emission and find overall good agreement, better than 50\% in all but one potentially problematic case (\citealt{Yusef-Zadeh2009}; see also \citealt{Koepferl2015}).

We estimate SFRs for the smaller scale structures in our catalog using two methods. The first method is to use the SFRs derived by \citet{Hatchfield2024}. \citet{Hatchfield2024} estimates SFRs in regions observed with the CMZoom Survey \citep{Battersby2020} by first combining archival data and catalogs of star-formation tracers, such as mid-IR point sources, methanol masers, UCHII regions,  to identify which of the 285 high-robustness CMZoom leaves from \citet{Hatchfield2020} is actively engaged in the star formation process. Each leaf is then given a designation of either `robustly' or `ambiguously' star forming as described in \citet{Hatchfield2024}. The mass of each star-forming leaf is divided by its free-fall time, multiplied by an estimated star formation efficiency \citep[SFE = 0.25 $\pm$ 0.15 is assumed in][]{Hatchfield2024} and corrected for the low-mass star formation to which the CMZoom survey is not sensitive \citep[see extensive details in][]{Hatchfield2024}. The total SFR of each cloud in CMZoom is then the summation of the SFRs of each star-forming leaf in the cloud. For this paper, we use the CMZoom SFRs for both robustly and ambiguously star-forming sources \citep[Table 3 in][]{Hatchfield2024}. We report the CMZoom SFR estimates for both the Herschel temperature calculation (usually around 20 K) as well as the CMZoom SFR estimate assuming a uniform temperature of 50 K. The former is the better assumption in a quiescent cloud, while the latter better captures the true mass, and therefore SFR for a warm, actively star-forming cloud. 

In order to utilize the SFRs calculated in \citet{Hatchfield2024}, we performed a manual comparison of each of our structures with the CMZoom clouds and found that only 12 regions had very good overlap and could be fairly compared side-by-side. For these 12 regions we include the CMZoom SFR estimates using both robustly and ambiguously star-forming sources. Since \citet{Hatchfield2024} involved a careful and thorough examination of multi-wavelength data of all of the CMZoom regions, 12 of which overlap with our hierarchical catalog, we consider this estimate of the SFR to be the most reliable. Therefore, when it is available, this estimate is used in the final column of Table \ref{tab:luminosity_properties} indicating it is the best available SFR. By default, we use the CMZoom SFR estimate using the Herschel temperature, however, for two highly active star-forming regions (cloud c: G0.380$+$0.050 and the 20\kms~cloud: G359.880-0.081), we use the 50 K CMZoom SFR estimate, as discussed in more detail at the end of this section. Both estimates are included in Table \ref{tab:luminosity_properties}.

For the second method of calculating SFRs in the smaller structures in our catalog, we follow the method presented in \citet{Barnes2017}. We summarize the method here, but refer the reader to that work for further details. Firstly, in order to calculate the total embedded stellar mass, we assume that the total IR luminosity for each cloud is a result of the reprocessed bolometric luminosity of the embedded stars within each cloud, which is then dominated by the single most massive star (i.e. M $\propto$ L$^x$, where x $\sim 1-3.5$; \citealp{Mould1982, Salaris2005}). 
This is a reasonable assumption in an actively star-forming cloud. In cases where external radiation is a significant contributor to the luminosity, our values set an upper limit of the possible SFR. 
To convert from a luminosity to the mass of the embedded object (M$_{*, {\rm tot}}$), we use the conversions from \citet{Davies2011}. As in \citet{Barnes2017}, we then extrapolate a total embedded stellar mass (M$_{*,{\rm tot}}$) from the most massive star in the system by solving for the normalisation $\beta$ in the following two equations:
\begin{equation}
    1 = \beta \int_{M_{*,{\rm max}}}^{\infty} m^{-\alpha} dm,
\end{equation}
where $\alpha$ = 2.3 and,
\begin{equation}
    M_{*,{\rm tot}} = \beta \int_{0.001}^{\infty} m^{1-\alpha} dm,
\end{equation}
where $\alpha = 0.3$ for $0.001<m/\Msun < 0.08$, $\alpha = 1.3$ for $0.08<m/\Msun<0.5$, and $\alpha = 2.3$ for $m/\Msun>0.5$ as in the Initial Mass Function (IMF) from \citet{Kroupa2001}.

Now that we have a total embedded stellar mass for each cloud, we need a timescale over which these stars are forming. \citet{Barnes2017} use the estimated time since peri-center passage for this timescale, which they assumed corresponded to the initiation of star formation in the cloud. By dividing the total embedded stellar mass by the time since star formation began, \citet{Barnes2017} estimated cloud-by-cloud SFRs. We instead use the calculated free-fall time for each cloud as the timescale for star formation. We calculate this free-fall time ($t_{\rm ff}$) using the structure mass ($M$), radius ($R$), and density ($n$) from Table \ref{tab:general_properties} and the standard free-fall time expression:
\begin{equation}
t_{ff} = \sqrt{\frac{3\pi}{32G\rho}},
 \end{equation}
where G is the gravitational constant and $\rho = M/\frac{4}{3}\pi R^3$.
We then divide the total embedded stellar mass (M$_{*, {\rm tot}}$) by the free-fall time to calculate the free-fall SFR (SFR FF) reported in Table \ref{tab:luminosity_properties}. The calculated free-fall times are also reported in Table \ref{tab:luminosity_properties}.

All of the SFRs are reported in Table \ref{tab:luminosity_properties}. The total IR estimate is considered valid in structures larger than 1000 pc$^2$, but otherwise is not reliable. For the overall dense CMZ (structure ID: 9), we find a total IR SFR estimate of 0.03 \Msun yr$^{-1}$. Unsurprisingly, this is lower than the estimate for the full inner region of the Galaxy from \citet{Barnes2017} of 0.07 \Msun yr$^{-1}$ because the area covered is substantially lower and is focused on dense gas, missing the bright radiation in the more diffuse ISM responsible for much of the total IR luminosity. When we compare the total IR SFR estimate against either of the two cloud-scale methods, the total IR SFR estimate vastly under-predicts the SFR in a single cloud. This is not surprising since it assumes that the luminosity comes from star formation spanning many spatial and time scales, which is not a valid assumption on the cloud scale.

The two cloud-by-cloud SFR estimates are considered to be reliable within leaves which generally represent individual molecular clouds. In general, the SFRs from the CMZoom and free-fall methods match each other quite well, however, this is not necessarily surprising since they both rely on the free-fall timescale in their calculations. A comparison of the twelve points with overlapping estimates from these two methods is shown in Figure \ref{fig:sfr_comparison}, with the CMZoom based SFR estimates from \citet{Hatchfield2024} on the y-axis and the free-fall based method derived in this work, based on the work from \citet{Barnes2017} on the x-axis. For each point, we include a 1-$sigma$ error bar. We compare the relationship between these SFR estimates and inspect their distance from the 1:1 line (dashed line). Most of the Herschel temperature points (our default assumption) are within 1-$sigma$ of the 1:1 line. However, there are two notable exceptions. There are two regions with significantly higher CMZoom SFRs than the free-fall estimate: cloud c (G0.380$+$0.050) and the 20\kms~cloud (G359.880-0.081). 
These clouds have CMZoom SFRs of 38.7 and 50.6 $\times 10^{-3}$ \Msun/yr respectively and free-fall SFRs of 3.3 and 12 $\times 10^{-3}$ \Msun/yr, a large discrepancy. 
These two regions are known to contain active star formation \citep[e.g.][]{Hatchfield2024, Lu2019a, Lu2019b}. Therefore, we select the CMZoom 50 K SFRs \citep[see][for details]{Hatchfield2024} as the better SFR esimate in Table \ref{tab:luminosity_properties} for these two clouds. Then, we have CMZoom SFRs of 10.2 and 9.2 $\times 10^{-3}$ \Msun/yr, respectively, showing much better agreement between our free-fall SFR estimates and the CMZoom 50 K SFR estimates for these two active star forming regions. There is also one low outlier (low CMZoom SFR compared with the free-fall method in this plot, which is the the Straw and Sticks clouds (ID: 38), for which we can offer no simple explanation except that these methods are imperfect and agreement within one sigma for the majority of points should perhaps instead be the surprise.
 
In the final column of Table \ref{tab:luminosity_properties} we report what we consider to be the best SFR estimate for each structure. For branches $>$ 1000 pc$^2$ in size this is the total IR SFR estimate. For leaves, this is the CMZoom-based estimate from \citet{Hatchfield2024} where available and the free-fall based estimate from this work in other cases. Based on the discussion in the previous paragraph, we use the 50 K CMZoom SFR estimate for two clouds, cloud c (G0.380$+$0.050) and the 20\kms~cloud (G359.880-0.081), and otherwise use the standard `Herschel tmperature' CMZoom SFR as the Best SFR. Branches that are smaller than 1000 pc$^2$ do not have a best SFR estimate. These best SFR estimates are used for the remainder of this work. 

There is substantial uncertainty in any measure of SFR, with many assumptions and caveats. In each of our leaves with a CMZoom or free-fall SFR estimate, we also have a total IR estimated SFR. The total IR estimated SFR is substantially lower than the CMZoom or free-fall SFR estimates. However, the total IR SFR relations were calibrated over large areas with stars in a range of evolutionary stages \citep[e.g.][]{Kennicutt2012} and are therefore not applicable to single molecular clouds bound by external pressure. When we focus in on the very densest gas, much of which is actively engaged in the star formation process already \citep[e.g.][]{Walker2021, Lu2021}, we would expect the SFR to be higher. The free-fall times reported in Table \ref{tab:luminosity_properties} of order 0.1 to 0.8 Myr are a reasonable estimate for expected timescales for high-mass star formation \citep{Battersby2017}. Focused future efforts on calibrating SFR prescriptions from galaxy to cloud scales is of great importance.

% Made in python using 'CMZ_diff_regions.ipynb'
%in folder ~/Dropbox/higal_cmz/nh2_pdfs/all_fits_files/
\begin{figure*}
\centering
\subfigure{
\includegraphics[width=1\textwidth]{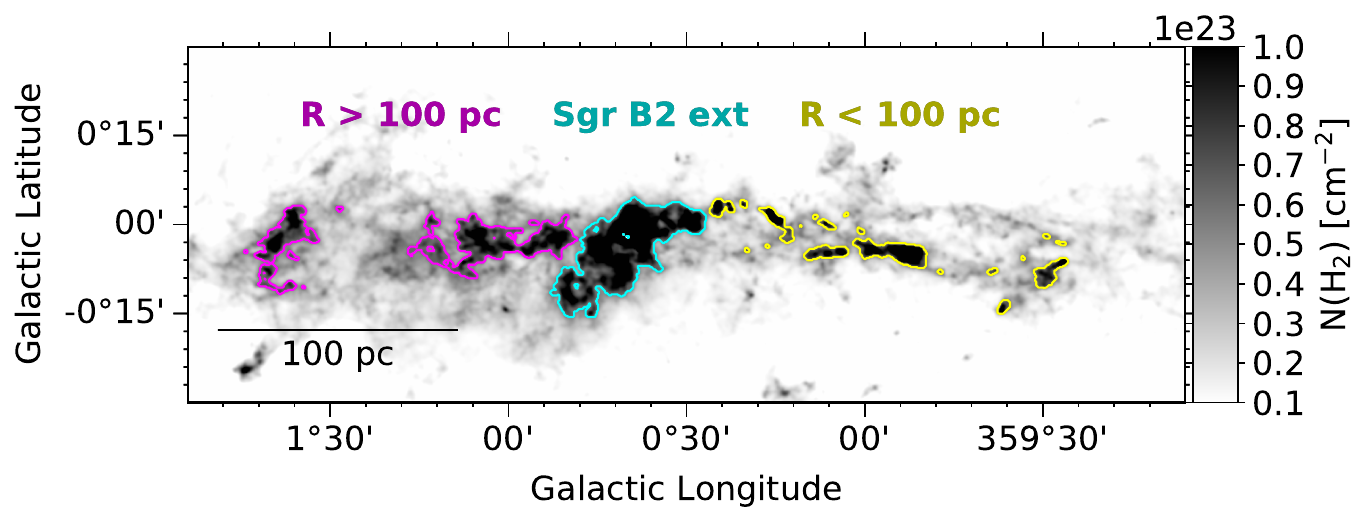}}
\subfigure{ 
\includegraphics[width=0.48\textwidth]{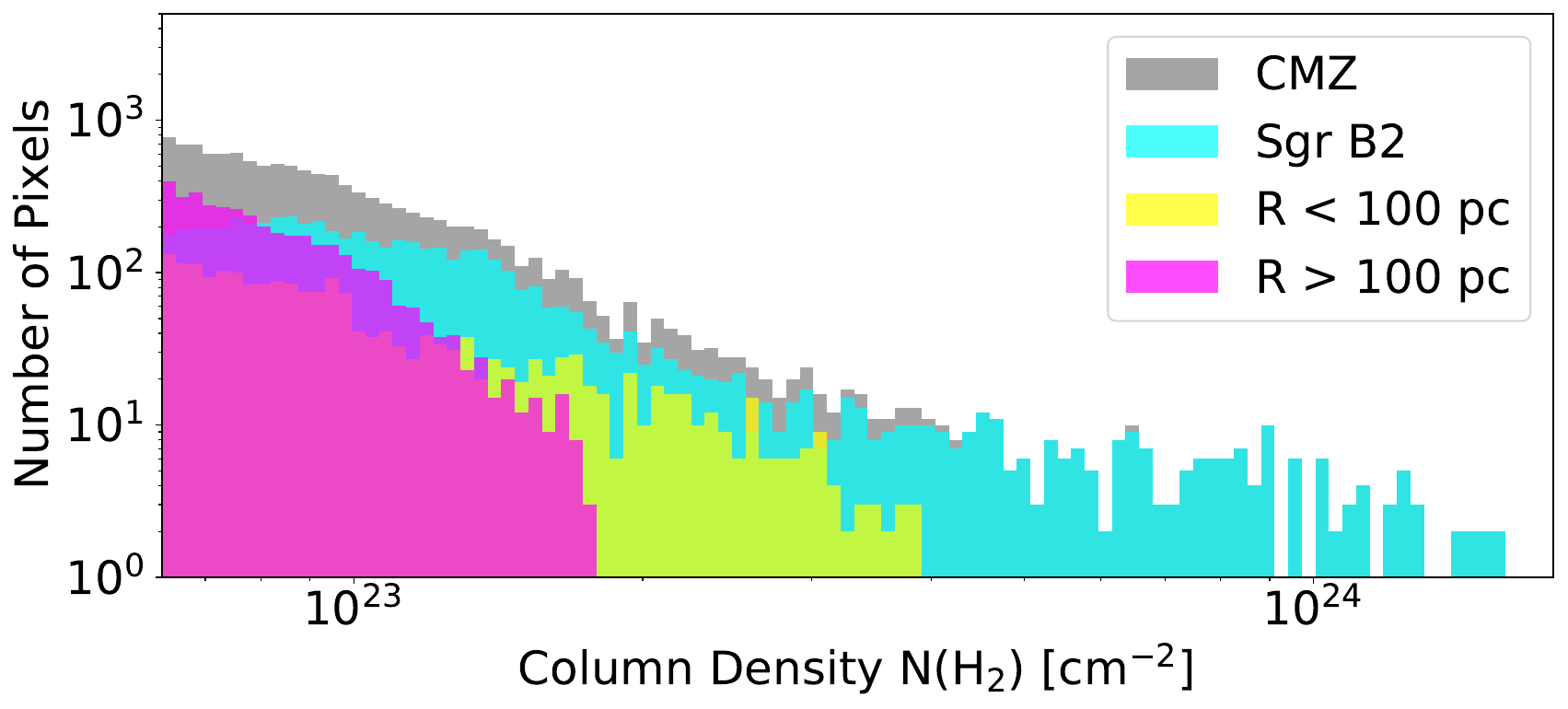}
\includegraphics[width=0.48\textwidth]{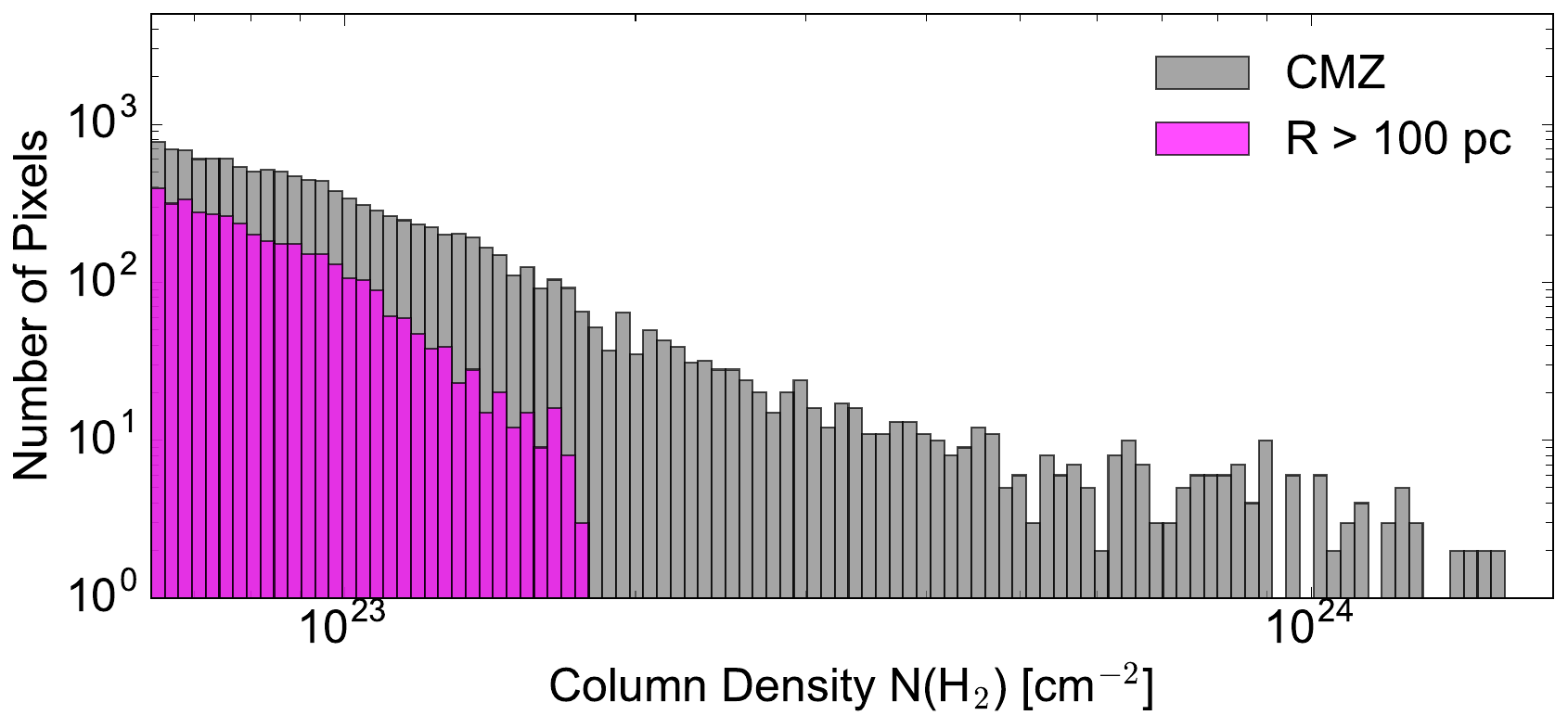}} \\
\subfigure{
\includegraphics[width=0.48\textwidth]{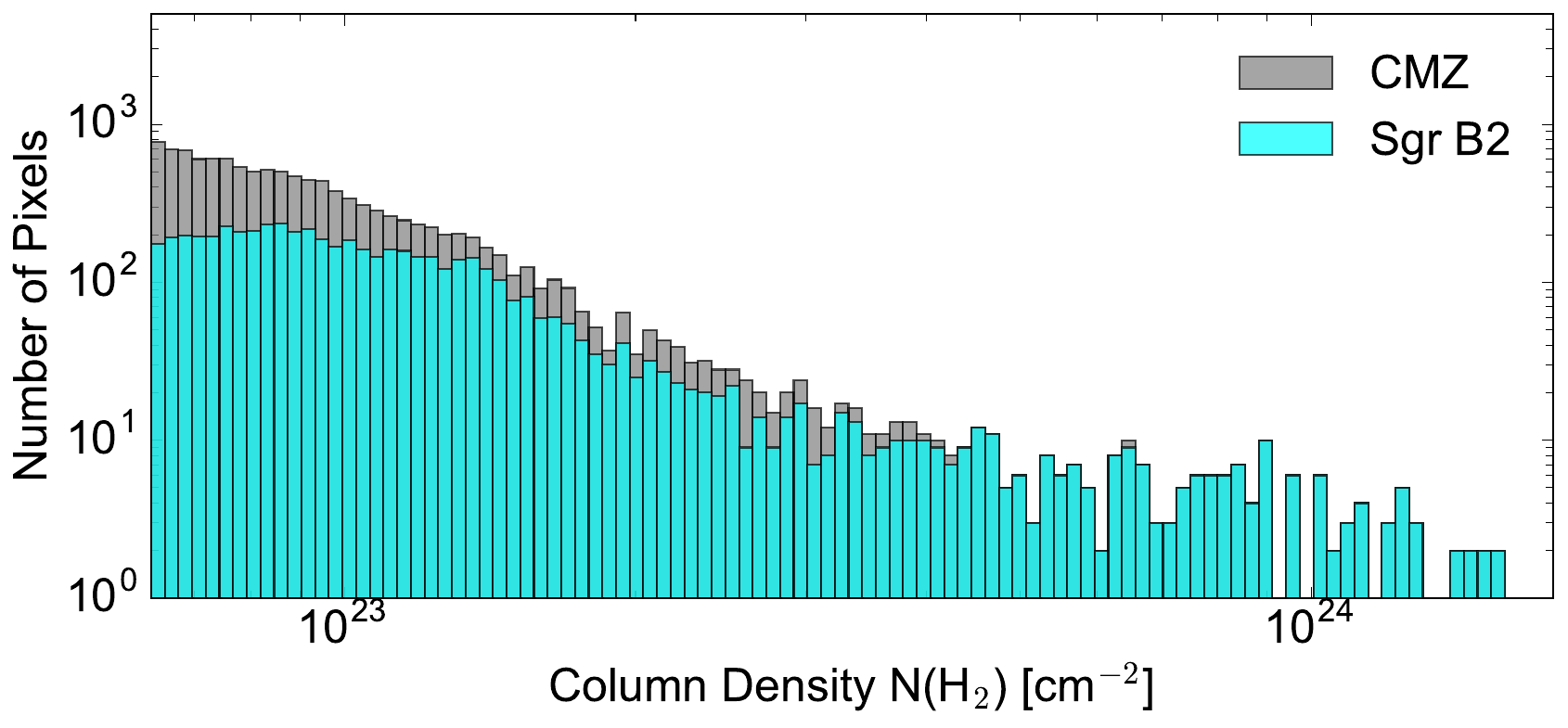}
\includegraphics[width=0.48\textwidth]{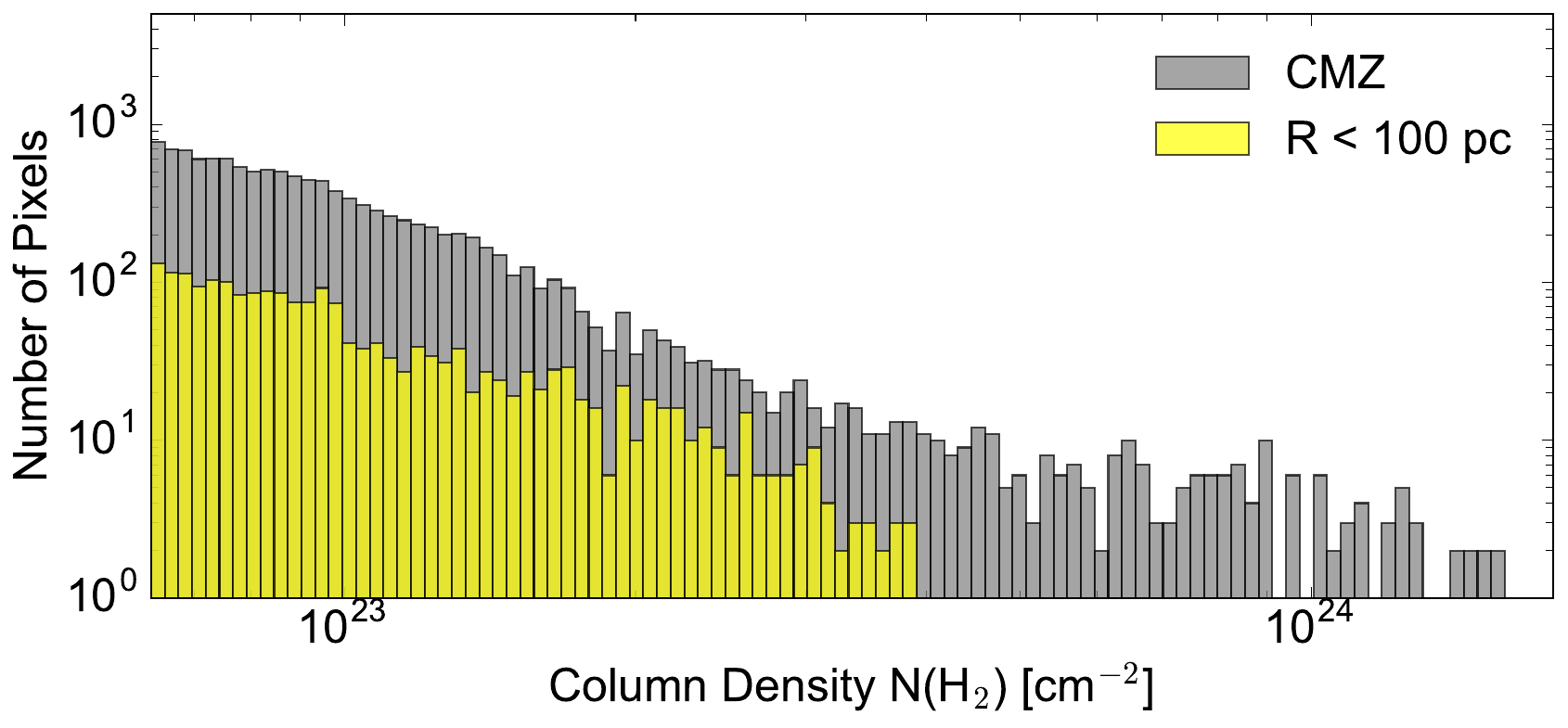}} \\
\caption{Column density PDFs for various regions in the CMZ highlight that most of the highest column density gas in the CMZ comes from the SgrB2 region, which has a very shallow slope, followed by the R $<$ 100 pc region, then by the R $>$ 100 pc region which has the highest fraction of low column density gas and steepest power-law slope. As described in Section \ref{sec:mle}, we used a minimum column density threshold of $N$(H$_2$) = 6 $\times$ 10$^{22}$ cm$^{-2}$ to separate the CMZ into these three key regions. The column density map is shown on \textit{top} and the three regions are highlighted in color contours. The colors correspond to the PDFs in the plot \textit{below}, while the gray shows the PDF over the entire CMZ above the same column density threshold. Each of these region PDFs is broken into select individual dendrogram structures in Figures \ref{fig:sgrb2_pdfs}, \ref{fig:inner100_pdfs}, \ref{fig:outer100_pdfs}.} 
\label{fig:fullcmz_pdfs}
\end{figure*}

% Updated Nov. 2024 with labels, using code: plot_dendro_PDFs_v2.ipynb (previously: NEW_plot_dendro_PDFs.ipynb
% SgrB2
\begin{figure*}[!htb]
\centering
\subfigure{
\includegraphics[width=0.48\textwidth]{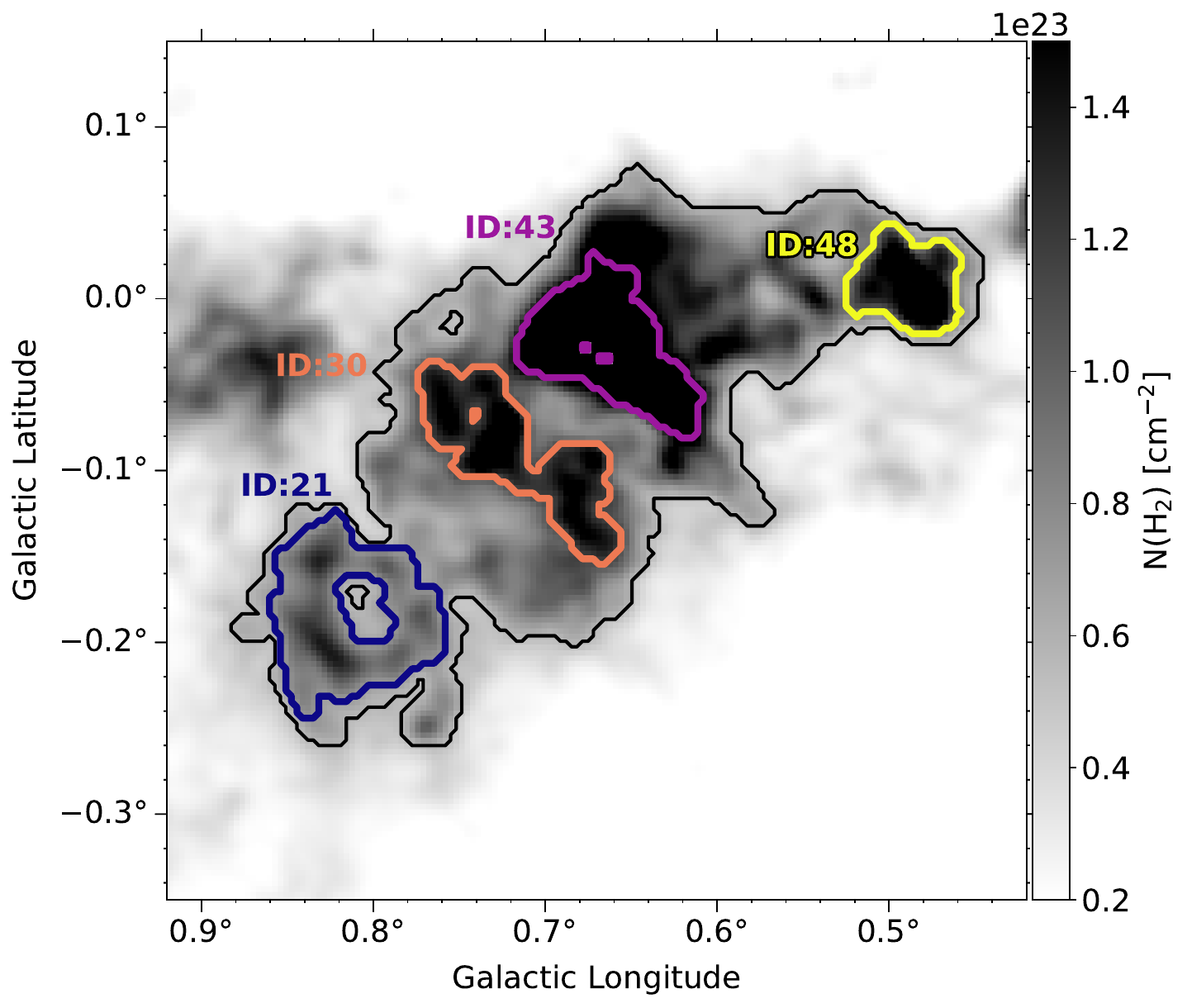}
\includegraphics[width=0.48\textwidth]{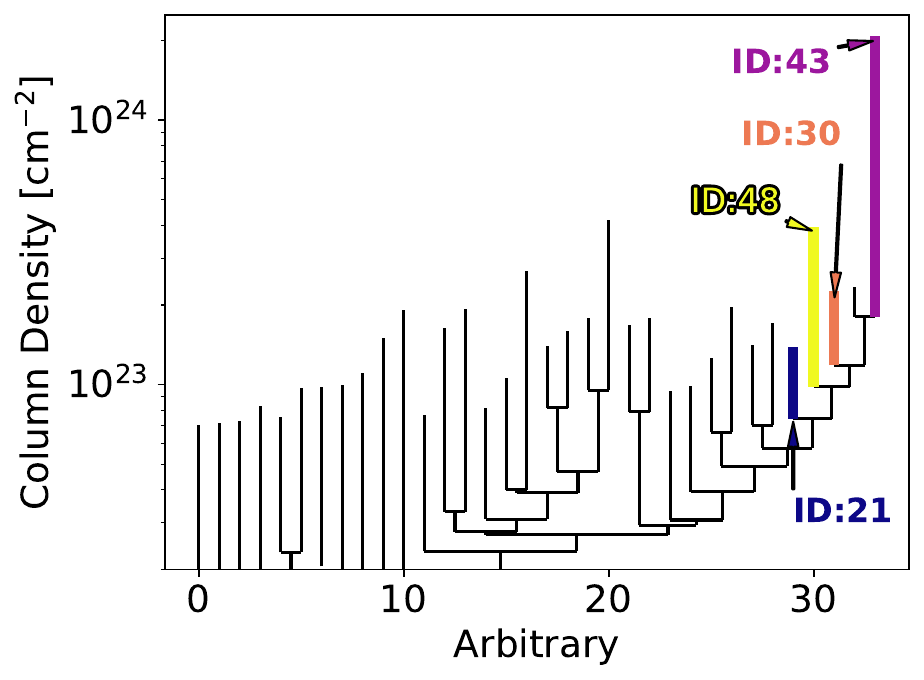}}
\subfigure{
\includegraphics[trim={0 55mm 0 55mm},clip, width=1\textwidth]{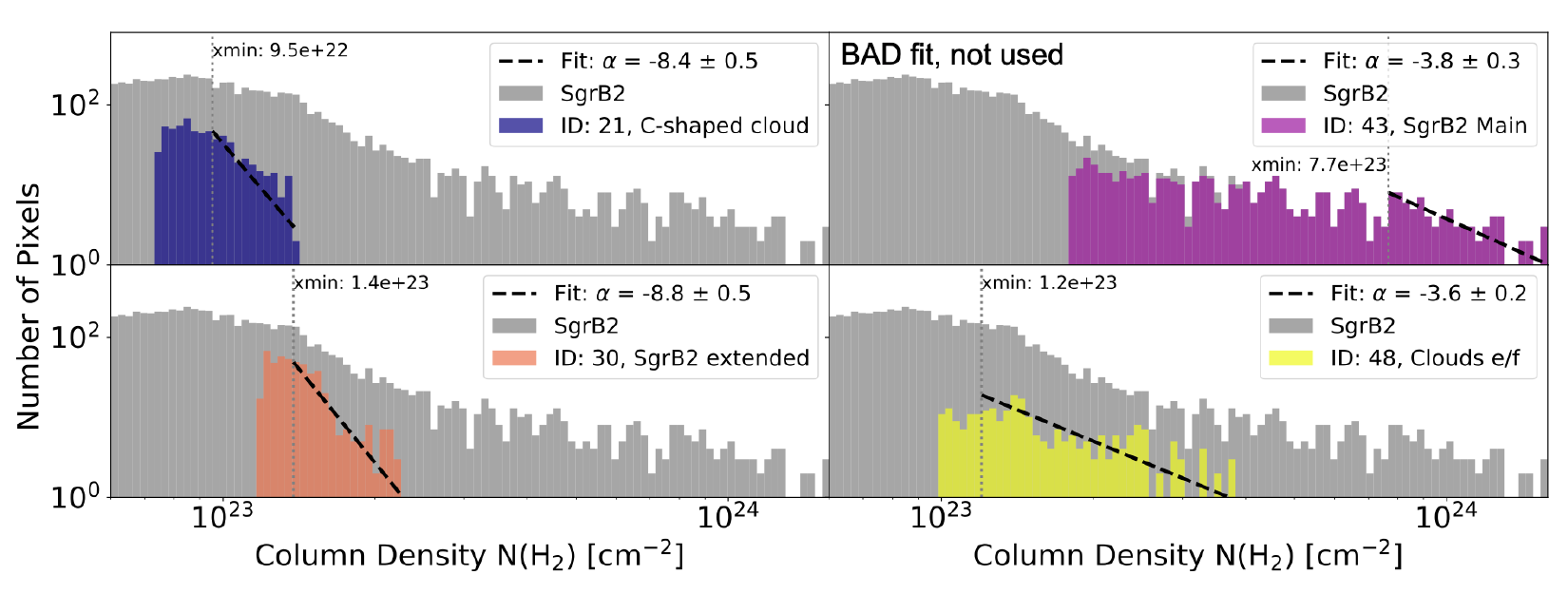}} 
\caption{The SgrB2 Main region comprises the highest column density gas in the CMZ. This figure highlights the extended SgrB2 region (black contour, same as the cyan contour in Figure \ref{fig:fullcmz_pdfs}) and four selected dendrogram structures within this region, color-coded and labeled in each plot: ID: 21 the `M0.8-0.2 ring' \citep{Nonhebel2024}, ID: 43 SgrB2 Main, ID: 30 SgrB2 extended, and ID: 48 Clouds e/f. Each dendrogram structure is highlighted as a contour in the image in the top left, as well as within the larger dendrogram structure tree in the top right in the same color. The bottom images show the column density histogram of the entire SgrB2 region (from Figure \ref{fig:fullcmz_pdfs}) in gray and the column density PDF of each selected structure in the same color in which the structure is highlighted in the upper panels. Power-law fits to the individual dendrogram structures are shown as a dashed black line, with the automatically calculated x$_{min}$ value marked with a vertical dotted gray line and the fit slope given in the legend. Fits to the overall region and individual dendrogram structures are also presented in Table \ref{tab:powerlaw} and discussed in Section \ref{sec:pdfs}. The fit to ID: 43, SgrB2 main was determined not to be representative of the power-law slope of the full region, so was excluded from further analysis (discussed in more detail in Section \ref{sec:pdfs}).}
\label{fig:sgrb2_pdfs}
\end{figure*}

% Updated Nov. 2024 with labels, using code: plot_dendro_PDFs_v2.ipynb (previously: NEW_plot_dendro_PDFs.ipynb
% Inner 100 pc
\begin{figure*}
\centering
\subfigure{
\includegraphics[width=0.6\textwidth]{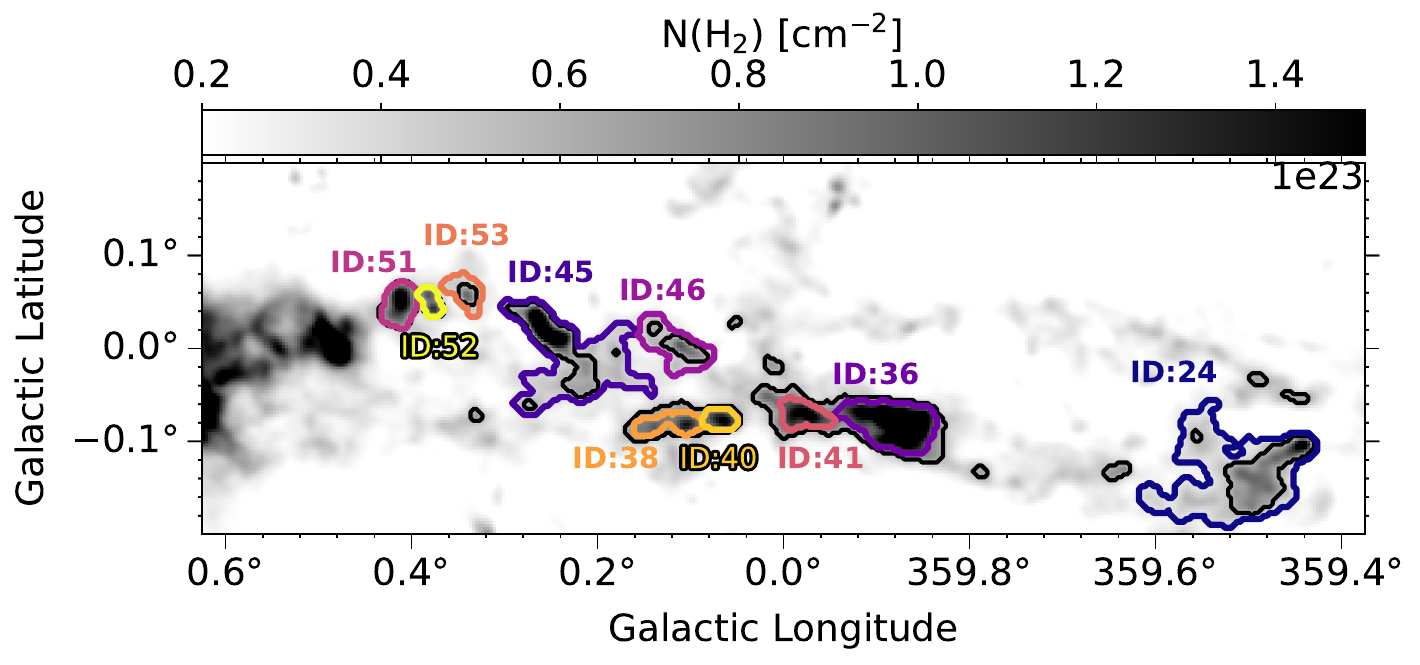}
\includegraphics[width=0.38\textwidth]{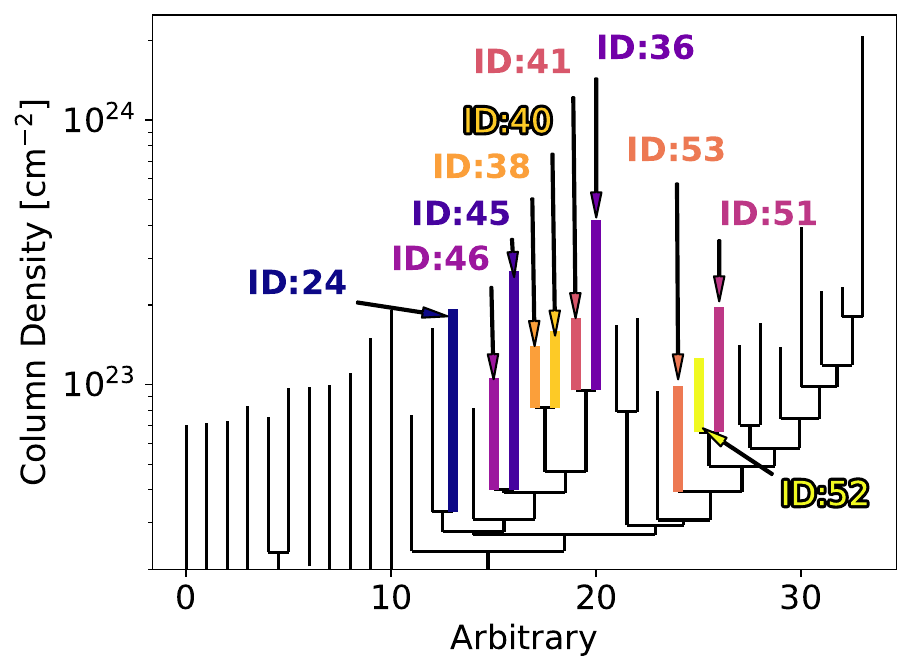}}
\subfigure{
\includegraphics[trim={0 0mm 0 0mm},clip, width=1\textwidth]{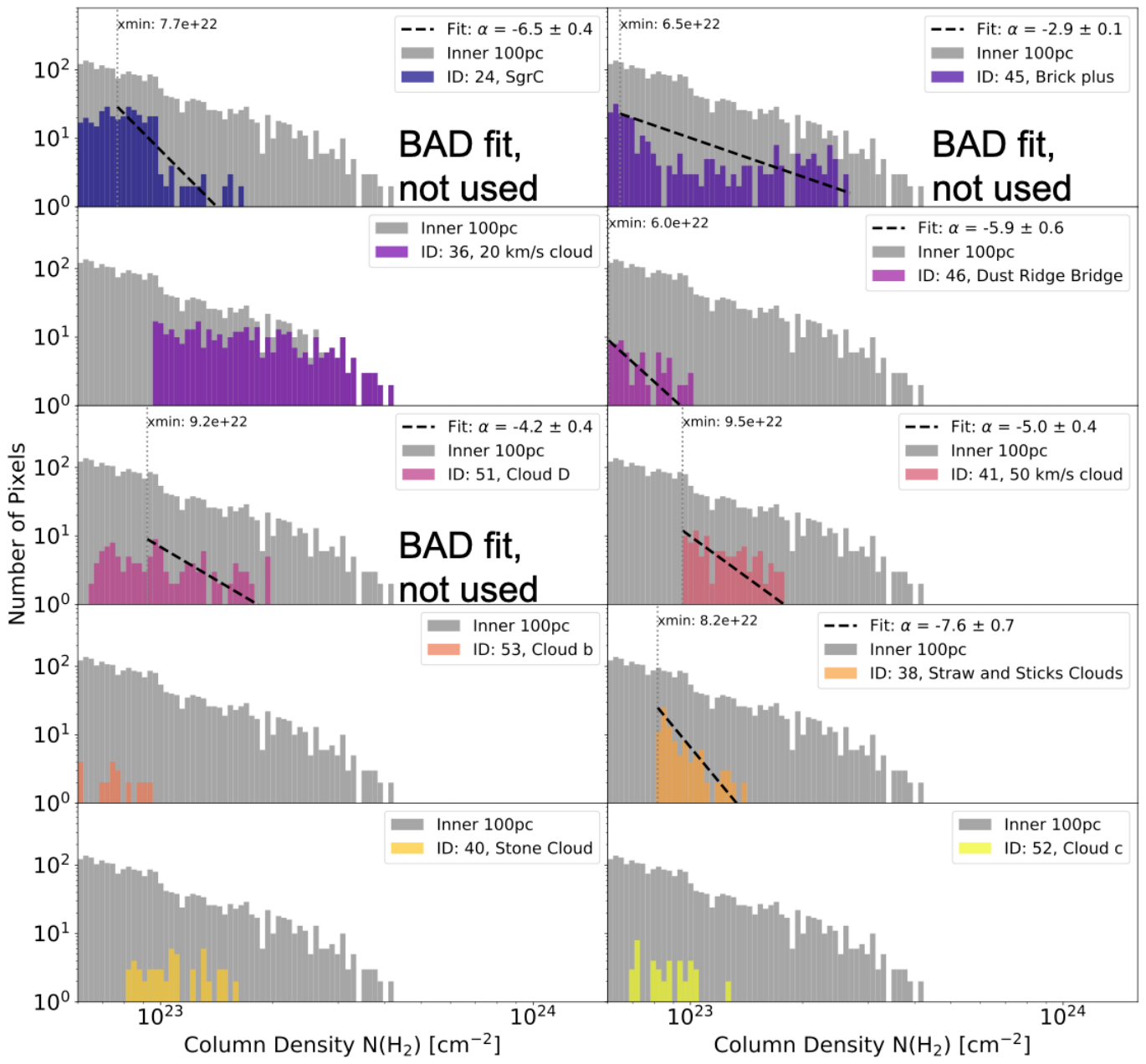}}
\vspace{-0.5cm}
\caption{The inner 100 pc region is not well-captured by a single column density contour, and is characterized by a series of high column density clouds scattered about the CMZ. This figure is the same as Figure \ref{fig:sgrb2_pdfs} but for structures in the inner 100 pc. The inner 100 pc region is shown with a black contour in the top left image as defined in Figure \ref{fig:fullcmz_pdfs} and ten individual dendrogram structures within this region, which are color-coded and labeled in each plot: ID: 24 SgrC, ID: 45 Brick plus extended, ID: 36 the 20\kms cloud, ID: 46 the dust ridge bridge, ID: 51 cloud D, ID: 41 the 50\kms cloud, ID: 53 cloud b, ID: 38 the straw and sticks clouds, ID: 40 the stone cloud, and ID: 52 cloud c. Fits that utilized fewer than 50 data points (structures 36, 53, 40, and 52) were excluded and the fits to structures 24, 45, and 51 were determined not to be representative of the power-law slope of the full region, so were excluded from further analysis (discussed in more detail in Section \ref{sec:pdfs}).}
\label{fig:inner100_pdfs}
\end{figure*}

% Updated Nov. 2024 with labels, using code: plot_dendro_PDFs_v2.ipynb (previously: NEW_plot_dendro_PDFs.ipynb
% Outer 100pc
\begin{figure*}[!htb]
\centering
\subfigure{
\includegraphics[width=0.6\textwidth]{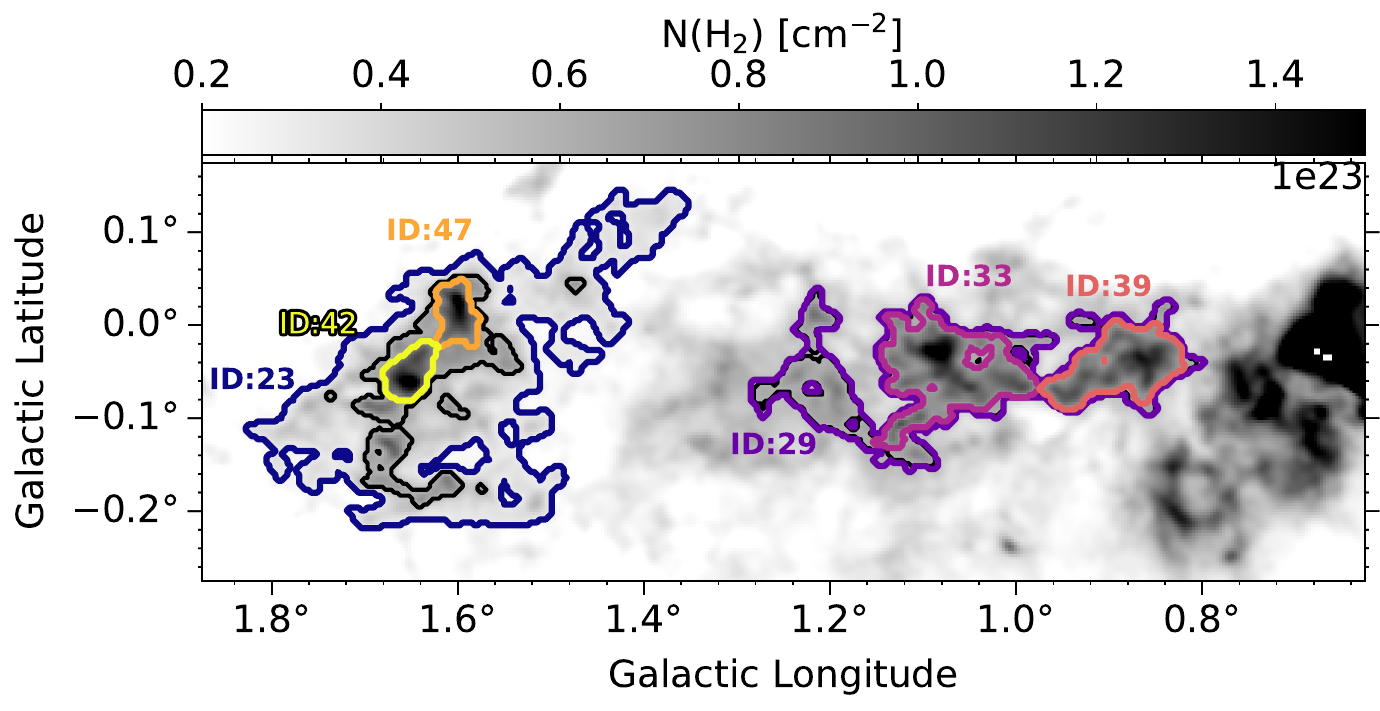}
\includegraphics[width=0.38\textwidth]{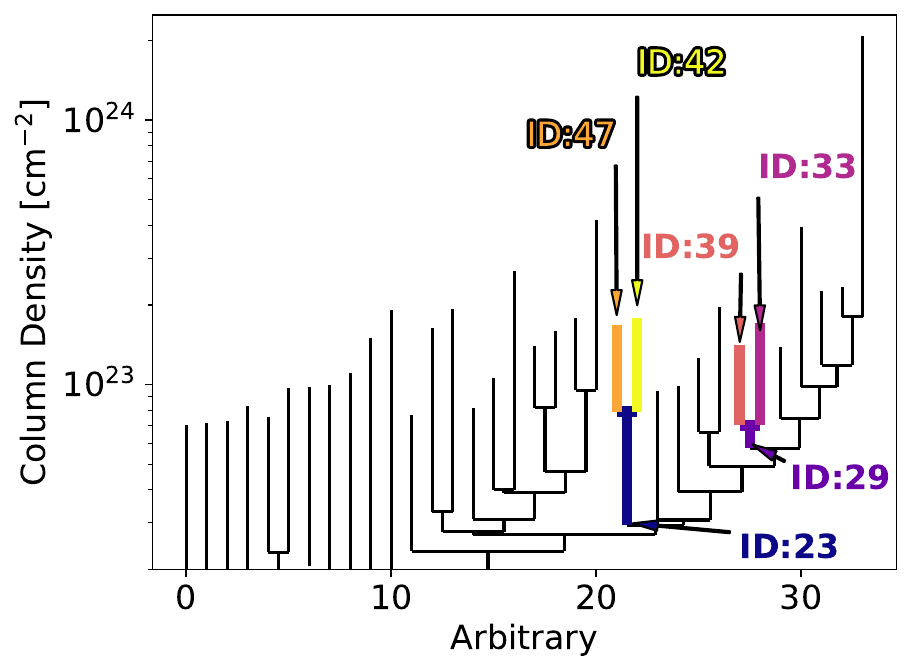}}
\subfigure{
\includegraphics[width=1\textwidth]{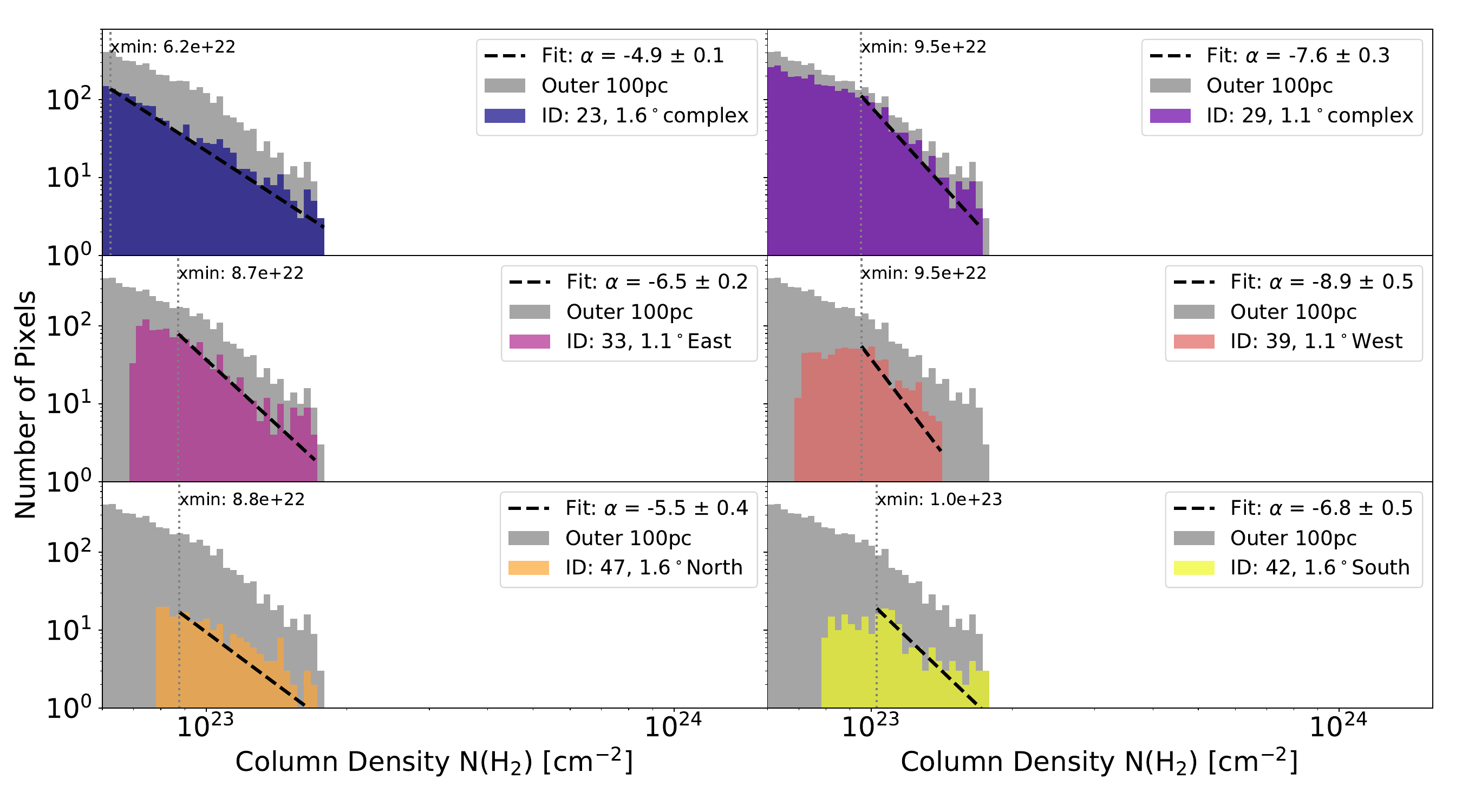}}
\caption{The outer 100 pc region contains two main complexes of clouds, the 1.6$^\circ$ and 1.1$^\circ$ complexes. This figure is the same as Figure \ref{fig:sgrb2_pdfs} but for structures in the outer 100 pc. This figure highlights the outer 100 pc region with a black contour as defined in Figure \ref{fig:fullcmz_pdfs} and six individual dendrogram structures within this region, which are color-coded and labeled in each plot: ID: 23 the 1.6$^\circ$ complex, ID: 29 the 1.1$^\circ$ complex, ID: 33 the 1.1$^\circ$ East cloud, ID: 39 the 1.1$^\circ$ West cloud, ID: 47 the 1.6$^\circ$ North cloud, and ID: 42 the 1.6$^\circ$ South cloud. The PDFs and power-law fits are discussed in more detail in Section \ref{sec:pdfs}.}
\label{fig:outer100_pdfs}
\end{figure*}

\begin{table*}
\centering
\caption{Power-law fits to the column density PDFs of various CMZ regions. We include here power-law fits to both the overall Sgr B2, inner 100 pc, and outer 100 pc regions highlighted in Figure \ref{fig:fullcmz_pdfs} as well as fits to individual dendrogram structures shown in Figures \ref{fig:sgrb2_pdfs}, \ref{fig:inner100_pdfs}, and \ref{fig:outer100_pdfs}. These are MLE fits where the lower-limit of the fit is automatically determined by the algorithm, and reported here as x$_{min}$. More details can be found in Section \ref{sec:mle}.}
\begin{tabular}{lclccccl}
\hline
 				& ID & Name 		& x$_{min}$ = $N$(H$_2$) & PL Slope & PL error	 & Best SFR estimate   & SFR method \\
 				&      &           		& [cm$^{-2}$]                    & $\alpha$ 	& 1 $\sigma$ 		  &  [M$_\odot$ / year]    &    \\
\hline \hline
\textbf{SgrB2} 		& 	& Overall 					& 1.7e+23 & -2.5 	& 0.1 	& --  		& -- \\
 				& 21 & M0.8-0.2 ring			& 9.5e+22 & -8.4 	& 0.5 	& 5.6e-3 	& free-fall   \\
 				& 43 & SgrB2 Main 				& --\footnote{Structures marked with a -- involved power-law fits to fewer than 50 data points or poor overall fits (Section \ref{sec:mle}) and are excluded.}  & --  & --  & 2.6e-2 & free-fall  \\
 				& 30 & SgrB2 extended 			& 1.4e+23 & -8.8  	&  0.5 	& 1.4e-2 	& CMZoom  \\ %$\pm$ 8.2e-3 
				& 48 & Clouds e/f 				& 1.2e+23 & -3.6  	&  0.2 	& 1.7e-2 	& CMZoom  \\ %$\pm$ 1.0e-2 
 \hline
\textbf{Inner 100 pc} 	&  	& Overall 					& 7.2e+22 & -3.3  	&  0.1 	& -- 		& -- \\
 				& 24 & SgrC 					& -- 		 & -- 		& -- 		& 7.7e-3 	& free-fall  \\ 
 				& 45 & Brick plus 				& -- 		 & -- 		& -- 		& 9.5e-3 	& free-fall  \\ 
 				& 36 & 20 km/s cloud 			& -- 		 & -- 		&  --		& 9.2e-3 	& CMZoom 50 K  \\ %$\pm$ 5.5e-3 
 				& 46 & Dust Ridge Bridge 			& 6.0e+22 & -5.9 	& 0.6 	& 1.7e-3 	& CMZoom  \\ %$\pm$ 1.0e-3
				& 51 & Cloud D 				& -- 		 & -- 		&  -- 		& 2.7e-3	& CMZoom  \\ %$\pm$1.6e-3
				& 41 & 50 km/s cloud 			& 9.5e+22 & -5.0  	&  0.4 	& 6.3e-3 	& CMZoom \\ %$\pm$3.8e-3
 				& 53 & Cloud b 					& -- 		 & -- 		& -- 		& 3.5e-3 	& free-fall  \\ 
 				& 38 & Straw and Sticks Clouds 	& 8.2e+22 & -7.6  	&  0.7 	& 4.0e-4 	& CMZoom   \\ %$\pm$ 3.0e-4 
 				& 40 & Stone Cloud 				& -- 		 & --  	& --		& 6.4e-3 	& CMZoom  \\ %$\pm$ 3.8e-3 
				& 52 & Cloud c 					& -- 		 & -- 		&  --		& 1.0e-2	& CMZoom 50 K  \\ %$\pm$6.1e-3  
 \hline
\textbf{Outer 100 pc}	& 	&  Overall  				& 1.0e+23 & -7.3  	&  0.2  	& -- 		& --  \\
 				& 23 & 1.6$^\circ$~complex 		& 6.2e+22 & -4.9  	&  0.1  	& 6.2e-4 	& total IR \\
 				& 29 & 1.1$^\circ$~complex 		& 9.5e+22 & -7.6  	&  0.3  	& -- 		& none  \\
 				& 33 & 1.1$^\circ$~East 			& 8.7e+22 & -6.5  	&  0.2  	& 7.1e-3 	& CMZoom  \\ %$\pm$ 4.2e-3
 				& 39 & 1.1$^\circ$~West 			& 9.5e+22 & -8.9  	&  0.5  	& 5.4e-3 	& free-fall \\
 				& 47 & 1.6$^\circ$~North 			& 8.8e+22 & -5.5  	&  0.4  	& 5.0e-3 	& CMZoom  \\ %$\pm$ 3.0e-3 
 				& 42 & 1.6$^\circ$~South 			& 1.0e+23 & -6.8 	&  0.5  	& 1.5e-3 	& CMZoom  \\ %$\pm$ 9.0e-4 
\hline
\end{tabular}
\label{tab:powerlaw}
\end{table*}

\subsection{Creating and Fitting Probability Distribution Functions}
\label{sec:mle}

Column density probability distribution functions (N-PDFs) provide a simplified view of the structure in a region that can be easily compared over different physical regions or between simulations and observations. Therefore, N-PDFs are widely used and discussed in the literature. We compute simple N-PDFs for the column density maps from Paper I and fit a simple power-law (details to follow), as well as perform comparisons with literature values. However, we caution the reader that there exists substantial complexity in the underlying N-PDF of any star-forming region \citep[e.g.][]{Chen2018} as well as in the observational extraction of a single number (e.g. a power-law slope) from these complex data \citep[e.g.][]{Alves2017}. 

Simulations of isothermal, supersonic turbulence predict a log-normal N-PDF \citep[e.g.][]{Ostriker2001, Padoan2011, Ballesteros2011, Hennebelle_review2012, Federrath2013} and when self-gravity is introduced, there exists a power-law tail at high densities \citep[e.g.][]{Klessen2000, Federrath2008, Kainulainen2009, Girichidis2014}. \citet{Burkhart2017} present an analytic expression for the expected transition point between the two functions. N-PDFs offer the tantalizing prospect of allowing for measurements of a number of physical parameters (such as turbulent driving and Mach number). Many observational studies have reported a log-normal N-PDF at low to moderate column densities with a possible power-law tail at high column densities, particularly in the presence of star formation \citep[e.g.][]{Schneider2022, Schneider2016, Rathborne2014, Schneider2013, Abreu-Vicente2015, Kainulainen2009}. However, work by \citet{Lombardi2015} and \citet{Alves2017} suggests that observed molecular cloud N-PDFs are consistent with power-laws and previous interpretations of a log-normal nature may be due to observational completeness biases \citep[note alternative view of][]{Ossenkopf-Okada2016}. 

In this work, we fit N-PDFs to data within closed column density contours to ensure completeness \citep{Alves2017}. We identify the lowest closed column density contour that allows for natural separation into key CMZ regions as $N$(H$_2$) $= 6 \times 10^{22}~\rm{cm}^{-2}$. Using this column density threshold, we divide the CMZ into three regions: R $<$ 100 pc (0.4\deg~ $\gtrsim$ $\ell$ $\gtrsim$ -0.6\deg), SgrB2 (0.8\deg~ $\gtrsim$ $\ell$ $\gtrsim$ 0.4\deg), and R $>$ 100 pc (1.8\deg~ $\gtrsim$ $\ell$ $\gtrsim$ 0.8\deg), with the exact regions shown as contours in Figure \ref{fig:fullcmz_pdfs}. These are the same regions as in Paper I, Section 4.2. The N-PDFs for the CMZ overall and for each region are shown in the bottom panels of Figure \ref{fig:fullcmz_pdfs}. 

We further investigate the nature of N-PDFs within each of these key CMZ regions by studying the N-PDFs of select individual dendrogram structures, which have already been identified as described in Section \ref{sec:dendrograms}. For consistency, we plot and fit these dendrogram structures above the same column density threshold as the larger regions, $N$(H$_2$) $= 6 \times 10^{22}~\rm{cm}^{-2}$, or their individual complete column density limit, whichever is higher. We present the three Figures \ref{fig:sgrb2_pdfs}, \ref{fig:inner100_pdfs}, and \ref{fig:outer100_pdfs} to show the location of each of these selected dendrogram structures on a map and within the dendrogram tree hierarchy. These figures also show the N-PDF of the overall CMZ region in which they are located and the N-PDFs of the individual sub-structures in the bottom panels. While there are a handful of saturated pixels in the Herschel map towards SgrB2, they are very few and do not affect the fits or analysis.

We choose to fit only the high column density portion of the N-PDFs, which is what we see above these complete closed column density contours. This portion of the N-PDF is generally considered to correspond to the star-forming gas in the cloud \citep[e.g.][]{Schneider2022, Schneider2016}. While there are many differing opinions on the choice of power-law vs.\ lognormal fits, we choose to fit simple power-laws to this high-column density gas, which allows for a simple, uniform fit to all of the data in our sample, and facilitates region inter-comparison. While lower column density material may require a lognormal function (or very steep power-law), fits to this material is more challenging to constrain observationally since it corresponds to dimmer emission, often near the noise limit of observations, and is often subject to incomplete column density cuts \citep[see e.g.][]{Alves2017}. For this work, we focus on simple power-law fits to the the high-column density N-PDFs in the CMZ.

We fit the high column density slopes of the N-PDFs using a Maximum Likelihood Estimator (MLE) method that produces a robust measurement of the slope that does not depend on choice of bin size. We fit a power law of the form: 
\begin{equation}
\label{eq:powerlaw}
p(N) = C N^{\alpha}
\end{equation}
where $p$ is the probability density, N is the column density, $\alpha$ is the power-law index slope, and $C$ is a constant. We use the python package {\sc powerlaw}\footnote{\href{https://github.com/jeffalstott/powerlaw}{https://github.com/jeffalstott/powerlaw}} from \citet{Alstott2014}, which relies upon statistical methods described in \citet{Clauset2009} and \citet{Klaus2011}, often used in astronomy \citep[e.g.][]{Veltchev2019}. This approach combines maximum-likelihood fitting methods and goodness-of-fit tests based on the Kolmogorov-Smirnov (KS) statistic and likelihood ratio. 

The {\sc powerlaw} package automatically determines an `x$_{\rm min}$' above which to fit the power-law slope and then fits a simple power-law (Equation \ref{eq:powerlaw}) above this value. In many cases, this x$_{\rm min}$ value is above the lowest closed column density contour, which means that the material below the x$_{\rm min}$ column density is not well-represented by the same power law, i.e. there is some truncation, either physical or observational, that causes a turnover. The {\sc powerlaw} algorithm accomplishes this by measuring the maximum-likelihood alpha parameter for each possible value of x$_{\rm min}$, then find the x$_{\rm min}$ that provides the closest match based on the KS test. The results of these fits are shown in Table \ref{tab:powerlaw} as well as in Figures \ref{fig:sgrb2_pdfs}, \ref{fig:inner100_pdfs}, and \ref{fig:outer100_pdfs}, and discussed in Section \ref{sec:pdfs}.

We exclude any fits performed using fewer than 50 data points (the 20\kms cloud, Cloud b, the Stone cloud, and Cloud c) or with fits determined to be poor upon visual inspection (SgrB2 Main, Sgr C, Brick plus, and Cloud d). The Stone cloud, cloud b, and cloud c simply did not contain enough data for a meaningful fit. In the case of the 20\kms~cloud, the highest column density points show a slightly steeper power law than the rest of the region, which flagged a truncation at a very high x$_{\rm min}$, with a power law fit of fewer than 50 data points and clearly not representative of the majority of the region. We also exclude the fits of four regions (SgrB2 Main, Sgr C, Brick plus, and Cloud d) that upon visual inspection do not appear to be high quality fits to the full N-PDF above the x$_{\rm min}$ value. For SgrB2 Main, the x$_{\rm min}$ value is 7.7 $\times 10^{23}$ cm$^{-2}$ and well above the majority of the pixels in the region, making it not representative of the full region. For Sgr C and Brick plus, the algorithm seemed to miss a clear truncation from a steep to more shallow power-law slope. Finally, for Cloud d, the number of data points about x$_{\rm min}$ is small and the fit simply does not seem to uniquely represent the population.

% Made by Dan Walker, will be added to github
\begin{figure*}
\centering
\includegraphics[width=0.7\textwidth]{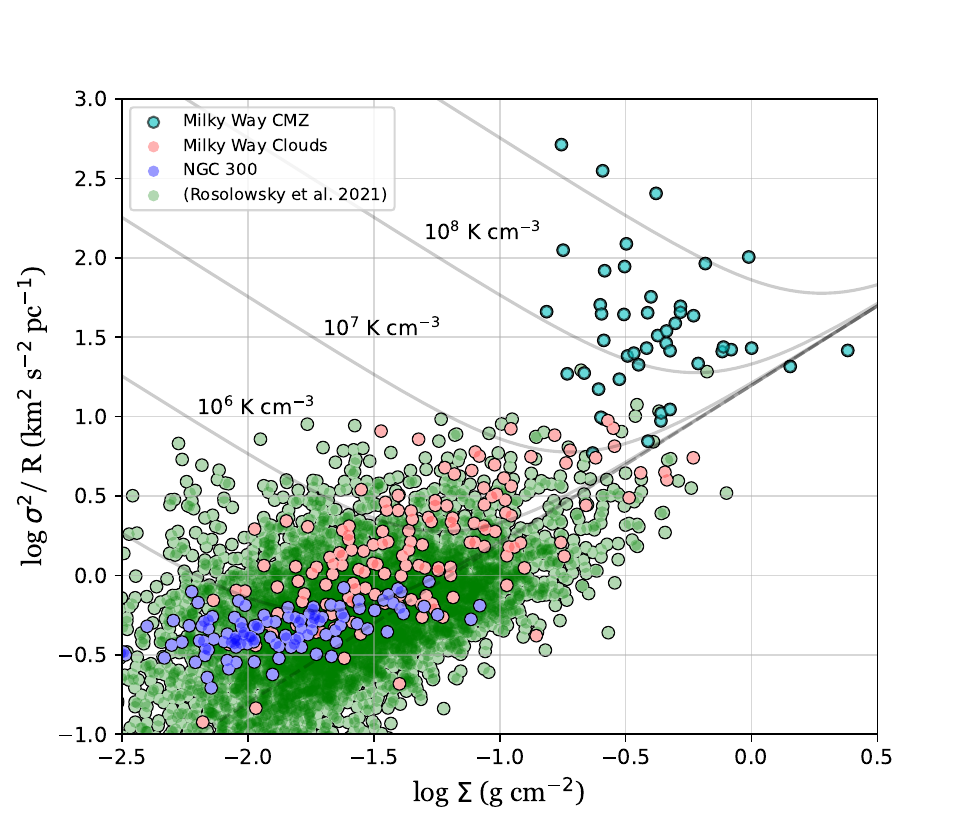}
\caption{Structures in the CMZ (boldly-outlined cyan points) have large linewidths, but they may be consistent with virial equilibrium in the presence of high external pressures (P$_{\rm e}/k$ $>$ 10$^{7}$ K cm$^{-3}$). 
Plotted here are velocity dispersions ($\sigma$) squared divided by the structure's effective radius vs. the surface density of gas for each structure ($\Sigma_{\rm gas}$). The cyan shows structures in the CMZ with data from Tables \ref{tab:general_properties} and \ref{tab:kinematic_properties}. In red are values for Milky Way clouds from \citet{Heyer2009}, in dark blue are clouds within NGC300 from \citet{Faesi2018}, and in green are clouds from the ten most nearby galaxies with PHANGS-ALMA from \citet{Rosolowsky2021}. See discussion in Section \ref{sec:size_lw}. Lines of virial equilibrium at constant external pressure are displayed as solid gray lines, using equation 11 from \citet{Field2011} starting with P$_{\rm e} = 10^5$ K cm$^{-3}$ and going up to P$_{\rm e} = 10^9$ K cm$^{-3}$. The dashed gray line represents simple virial equilibrium with no external pressure. Points below a gray curve would be in virial equilibrium assuming the external pressure of the plotted curve. This figure is inspired by \citet{Field2011}, \citet{Walker2018}, and \citet{Krieger2020}, and it highlights the complications of determining virial equilibrium in the presence of external pressure. 
}
\label{fig:size_lw}
\end{figure*}

\section{Discussion}
\label{sec:discussion}

\subsection{Comparison of Size, Linewidth, and Surface Density with other Regions}
\label{sec:size_lw}

In Figure \ref{fig:size_lw} we compare the sizes, linewidths ($\frac{\sigma^2}{R}$), and gas surface densities ($\Sigma$) of CMZ structures identified in this work (in boldly-outlined cyan) with Milky Way clouds from \citet{Heyer2009} (in red), molecular clouds in NGC300 from \citet{Faesi2018} (in dark blue), and a sample of molecular clouds towards ten of the nearest galaxies in the PHANGS-ALMA survey \citep{Rosolowsky2021}. The PHANGS-ALMA preliminary catalog from \citet{Rosolowsky2021} has a common 90 pc resolution, \citet{Faesi2018} achieves a spatial resolution of 10 pc, and \citet{Heyer2009}'s resolution ranges from about 0.3 to 3.0 pc at different molecular cloud distances. Our observations have a resolution of 1.5 pc, and the largest structures in the catalog are about 120 pc in size. Together, these data sample both galaxy disks and centers with physical scales that overlap. Our Milky Way CMZ probes higher surface densities than other samples for two reasons: 1) the surface density in the CMZ is higher than the Galactic disk, 2) compared with extragalactic CMZ's from \citet{Rosolowsky2021}, we have a higher physical resolution and therefore, when observing dense gas, can probe higher average densities. 

Figure \ref{fig:size_lw} is inspired by an investigation into the effects of external pressure on virial equilibrium from \citet{Field2011}. We use equation 11 from \citet{Field2011} to plot the solutions of pressure-bounded virial equilibrium for given external pressures in Figure \ref{fig:size_lw} from P$_{\rm e}/k$ $=$ 10$^{5}$ K cm$^{-3}$ to P$_{\rm e}/k$ $=$ 10$^{9}$ K cm$^{-3}$, the same equation as used in other recent works \citep[e.g.][]{Krieger2020, Callanan2023, Walker2018}, see also \citet{Myers2022}:

\begin{equation}
\label{eq:pressure_eq}
    \frac{\sigma^2}{R} = \frac{1}{3}\left(\pi\Gamma G\Sigma + \frac{4P_e}{\Sigma}\right)
\end{equation}

where $\Gamma$ is assumed to be 0.73 \citep[see][]{Field2011}. Of the 56 CMZ structures in Table \ref{tab:general_properties}, 11 do not have linewidth measurements from the MOPRA data (see Section \ref{sec:kinematic_properties}) and are therefore excluded (IDs 0-7, 10, 55, and 56).

In general, the CMZ shows larger $\frac{\sigma^2}{\rm{R}}$ and $\Sigma$ values than Galactic disk measurements from \citet{Heyer2009} in the Milky Way or \citet{Faesi2018} in NGC300. Our measurements probe higher densities than the sample from \citet{Rosolowsky2021} that also includes galaxy centers because our higher physical resolution (1.5 pc vs. 90 pc) means that our selected structures are focused on gas of higher average density, but we agree with their conclusion that galaxy centers are on average higher than galaxy disks (and farther from virial equilibrium) in this plot. CMZ structures are farther above the dashed gray line which shows simple virial equilibrium in the absence of external pressure. Several authors argue that the majority of gas in the CMZ is unbound \citep[e.g.][]{Callanan2023, Myers2022, Lu2019b}. \citet{Krieger2020} performs a detailed apples-to-apples comparison the CMZ with the center of NGC253 and find that many structures in the CMZ on scales of $\sim$ 10-100 pc scales follow the expectations for gravitationally bound objects with virial parameters of about 2-3, while others have high virial parameters. Figure \ref{fig:size_lw} shows that CMZ structures from this work are largely inconsistent with simple virial equilibrium, but could be considered to be in pressure-bounded virial equilibirum if in the presence of external pressures exceeding P$_{\rm e}/k$ $>$ 10$^{7-9}$ K cm$^{-3}$. As noted previously, our linewidths are from a simple moment analysis, as in \citet{Heyer2009, Faesi2018, Rosolowsky2021}. However, multiple structures along the line of sight complicate the interpretation, and we refer the reader to Paper III for details on the kinematics of clouds within our catalog as well as \citet{Shetty2012, Kauffmann2017a, Henshaw2016b} for detailed kinematics of the CMZ overall. 

Figure \ref{fig:size_lw} demonstrates that structures in the CMZ are either largely unbound or require a high external pressure ($P_{\rm e} / k \sim 10^{7-9}$ K cm$^{-3}$) to be considered in pressure-bounded virial equilibrium, much higher than clouds in NGC 300 or throughout the rest of the Milky Way. This high external pressure is consistent with large-scale turbulent pressure estimates in this environment \citep[e.g.][]{Walker2018, Rathborne2014, Kruijssen2013}. 

As we have the full hierarchy of the structures from our analysis, we can assess whether the pressure within a parent structure is sufficient to confine the embedded leaf structure. Under the na{\"i}ve assumption that the internal pressure of a branch structure is equivalent to the pressure exerted upon the leaf, we estimate the total external pressure as the sum of the turbulent, gravitational, and thermal pressures using

\begin{equation}
\label{eq:pturb}
    P_{turb} = \frac{1}{3} \rho \sigma^2
\end{equation}

\begin{equation}
\label{eq:pgrav}
    P_{grav} = \frac{3}{4\pi} \frac{GM^2}{R^4}
\end{equation}

\begin{equation}
\label{eq:ptherm}
    P_{therm} = nk_BT
\end{equation}

where all observed quantities (except for temperature) are the same as those reported in Tables \ref{tab:general_properties} and \ref{tab:kinematic_properties}. For the gas temperature, we assume a single value of 50~K in all cases. While this is a significant oversimplification, it is in line with the CMZ average of 50-200~K, or $\sim$50 K in the densest regions \citep[e.g.][]{Ginsburg2016, Krieger2017}. Moreover, this thermal pressure is orders of magnitude lower than the turbulent and gravitational pressures, and so any change in the temperature assumption has a negligible impact on the total pressure. We assume isotropic turbulence and a sphere of uniform density when calculating the turbulent and gravitational pressures. While these assumptions are highly simplistic, a detailed study is beyond the scope of this work, and these results serve as a rough estimate of the internal pressures, and whether or not they are of the same order of magnitude as that required for pressure-bounded virial equilibrium .

We then estimate the total pressure in each parent branch structure as $P_{\textrm{tot}}$ = ($P_{\textrm{turb}}$ + $P_{\textrm{grav}}$ + $P_{\textrm{therm}}$), and compare this to the external pressure required to confine the leaf within, which is estimated by rearranging Equation \ref{eq:pressure_eq} for $P_{\textrm{e}}$. The $P_{\textrm{e}}$ values for each leaf and the $P_{\textrm{tot}}$ of the corresponding parent branch are given in Table \ref{tab:luminosity_properties}, along with the ratio of $P_{\textrm{tot}}$/$P_{\textrm{e}}$, where values \textgreater\ 1 indicate pressure confinement. Overall, we find that the majority of the leaves are consistent with being pressure-confined by their parent branch, and almost all are consistent within a factor of $\sim$ 2. While there are significant caveats related to the assumptions in this analysis, the results illustrate that the pressures involved are broadly sufficient to confine the structures.

Another possible source of pressure comes from the large-scale hot gas \citep[e.g.][]{Blitz1993}. For example, \citet{Oka2019} argues for a large filling fraction of warm, low density gas which could provide the required pressure. We would need to invoke hot (T$>10^6$ K) gas in order to produce this pressure thermally, if we assume that the required density is lower than or comparable to that of our molecular clouds. Understanding the magnitude, nature and origin of  pressure in the CMZ is critical for understanding how clouds in the CMZ become bound and form stars. \citet{Kauffmann2017a} uncovers an unusually steep size-linewidth relationship in the CMZ and finds that on small scales the velocity dispersion goes down. 

% Made by Danya Alboslani, code on github
\begin{figure*}
\subfigure{
\includegraphics[trim = 0.5cm 0.5cm 1cm 1cm, clip, width=0.50\textwidth]{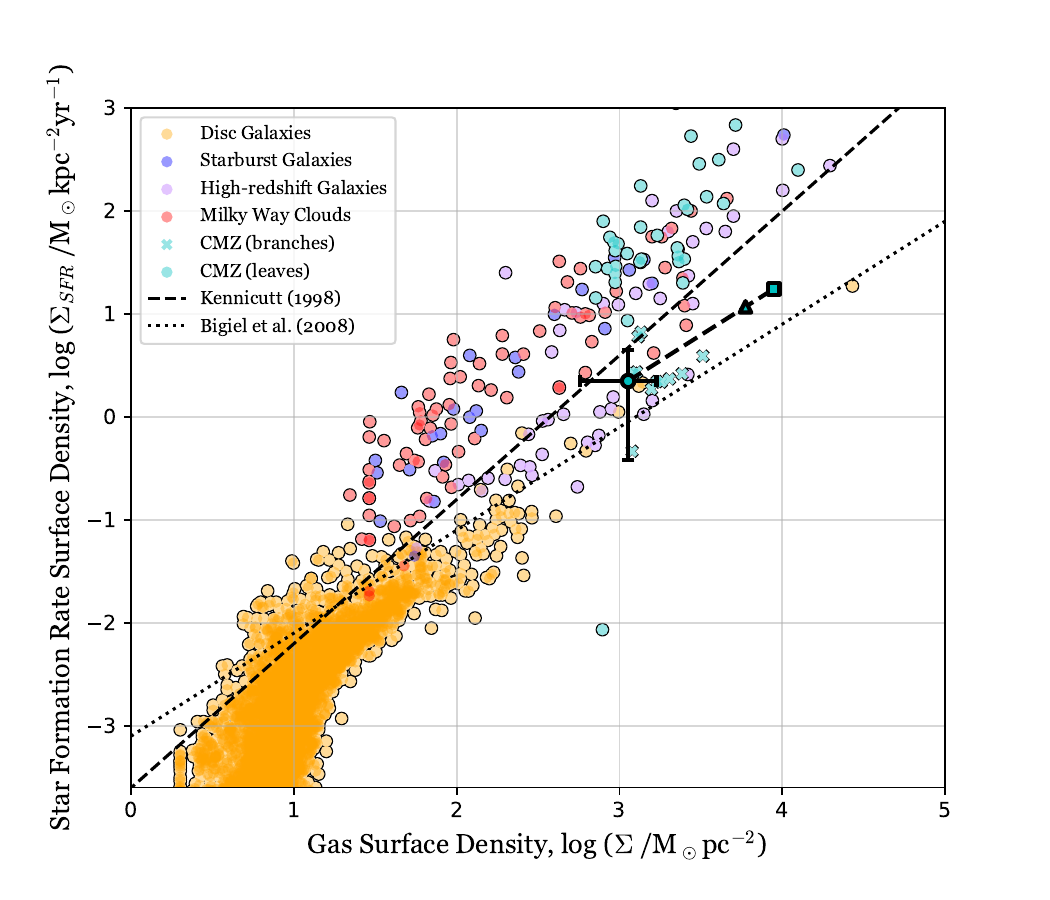}
\includegraphics[trim = 0.5cm 0.5cm 1cm 1cm, clip,width=0.50\textwidth]{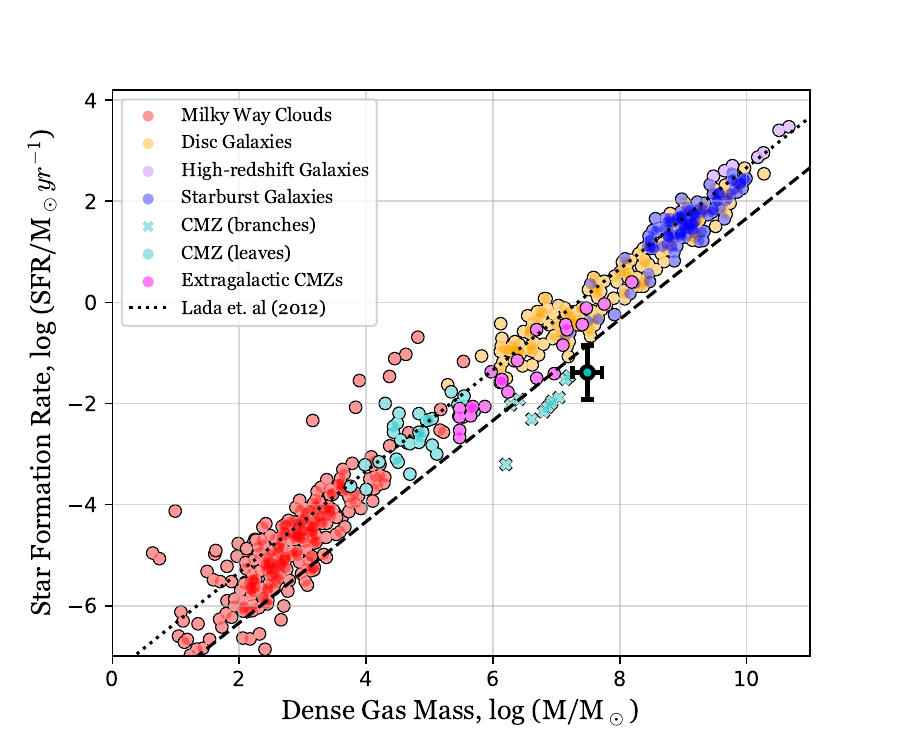}}
\caption
{We compare the SFR vs. density properties of CMZ structures identified in this work (cyan) with structures in the literature, from Milky Way clouds to high-redshift galaxies. While the CMZ overall (bold cyan circle) falls an order of magnitude below the dense gas star formation relation, the individual clouds in the CMZ show better agreement. In both plots, the cyan points are CMZ structures identified in this work (circles are molecular-cloud-sized leaves and crosses are larger branches) with values from Tables \ref{tab:general_properties} and \ref{tab:luminosity_properties}, while the boldly outlined cyan shapes (circle, triangle, and square) are the CMZ overall from \citet{Henshaw2023} \citep[each assuming a different CMZ geometry in the density calculation, see Section 3.1.3 in][for details]{Henshaw2023}. The gray-outlined colored points are from the literature \citep[see][and references therein]{Henshaw2023, Jimenez-Donaire2019, Krumholz2014}
\textit{Left:} The SFR surface density ($\Sigma_{\rm{SFR}}$) is plotted as a function of the gas surface density ($\Sigma_{\rm{gas}}$) with the scaling relations from \citet{Kennicutt1998} and \citet{Bigiel2008} (power-law slopes of n=1.4 and n=1.0, respectively) plotted on top. \textit{Right:} The SFR vs. mass of dense gas ($M_{\rm dense}$) with the scaling relation from \citet{Lada2012} shown as a dotted line and a factor of 10 below this relation shown as a dashed line.\textsuperscript{a}
}
\small\textsuperscript{a}{The out-lying low cyan point on the left plot is G357.070-0.770 (Structure ID: 1), which is an isolated structure, far off the Galactic plane (b = -0.770), and is likely not in the CMZ.}
\label{fig:exgal}
\end{figure*}
%\small\textsuperscript{a}{The out-lying low cyan point on the left plot is G357.070-0.770 (Structure ID: 1), which is an isolated structure, far off the Galactic plane (b = -0.770), and is likely not in the CMZ.}

\subsection{Comparison of Global SFRs with Other Regions}
\label{sec:exgal}

In Figure \ref{fig:exgal} we compare the star formation rates and physical properties of CMZ structures derived in this work with other regions in the literature, from Milky Way clouds to high-redshift galaxies. These plots are largely built on the review paper from \citet{Henshaw2023} and pull data and relations from \citet{Krumholz2014}, \citet{Jimenez-Donaire2019}, \citet{Kennicutt1998}, \citet{Bigiel2008}, \citet{Lada2012}, \citet{Querejeta2019}, \citet{Jiang2020}, \citet{Beslic2021} and references therein. In both plots the cyan points are CMZ structures from this work (circles are molecular-cloud-sized leaves and crosses are large-scale branches), with data from Tables \ref{tab:general_properties} and \ref{tab:luminosity_properties}. The boldly outlined cyan points are overall CMZ points from \citet{Henshaw2023}. 

The methods for estimating both star formation rate and gas mass vary between and within the samples presented. Without doubt, some of the scatter observed in these plots could be attributed to the intrinsic uncertainty in any particular method for estimating these quantities. We do not presently have well-established SFR calibrations on molecular cloud scale structures (leaves). In Section \ref{sec:sfr}, we argue that our approach for cloud-scale SFRs is a reasonable first attempt and is based on free-fall timescales of about 0.1 - 0.8 Myr (Table \ref{tab:luminosity_properties}) and is consistent with expectations for high-mass star formation \citep[e.g.][]{Battersby2017}. Our CMZ leaf points align with the global SFR relationships with no particular tuning on our part, just SFR estimates derived from first astrophysical principles. However, these cloud-scale (leaf) SFRs are still uncertain and more detailed future work calibrating SFRs on cloud-scales would be highly valuable.

To estimate the global uncertainties in this plot, we can refer to our own CMz, where the scatter in the SFR measurements depending on technique account for about a factor of two of the variation observed \citep{Barnes2017}. Without a more detailed apples-to-apples comparison of measurement techniques in these samples, we cannot be sure how much of the scatter or trends may be due to variation in measurement techniques, however, if the global CMZ SFR measurement is to be any guide, the measurement variation of about a factor of two is much less than the observed scatter of more than 1 dex.

In the left plot, we compare the star formation rate surface density ($\Sigma_{\rm{SFR}}$) with the gas surface density ($\Sigma_{\rm{gas}}$). \citet{Henshaw2023} compare the CMZ data point overall (cyan with bold outline, with three possible positions in this plot depending on geometrical assumption - circle assumes face-on disk geometry, triangle a face-on ring geometry, and square an edge-on geometry) and find that the CMZ overall is roughly consistent, or marginally inconsistent depending on geometry, with the \citet{Kennicutt1998} relationship. \citet{Kruijssen2014} report that the CMZ overall is consistent with the \citet{Bigiel2008} relationship which would support the hypothesis that some mechanism is inhibiting star formation in the CMZ. 

We find considerable scatter in the large sample of CMZ structures compiled in this work, however, the majority of the points lie above the \citet{Kennicutt1998} relationship and closely follow comparable regions from the Milky Way, starburst, and high-redshift galaxies. CMZ data points associated with leaves (circles in Fig. \ref{fig:exgal}) tend to represent molecular cloud sized structures and follow this trend most closely. 
CMZ data points associated with branches (crosses in Fig. \ref{fig:exgal}) are more representative of the CMZ overall, including gas not associated with individual molecular clouds, and lie between the \citet{Kennicutt1998} and \citet{Bigiel2008} relationships, in close agreement with the CMZ overall data points from \citet{Henshaw2023}.

The right plot of Figure \ref{fig:exgal} compares the various regions' SFRs with their dense gas mass. As discussed in detail in \citet{Henshaw2023}, the CMZ overall (boldly-outlined cyan circle) is clearly inconsistent with this so-called dense gas star formation relationship \citep[see, e.g.][]{Longmore2013a}, first proposed by \citet{Gao2004} and extended by \citet{Lada2010, Lada2012}. 
However, our new detailed analysis finds that while branches (crosses in Fig. \ref{fig:exgal}) similarly show a significant deviation from the dense-gas star formation relation, leaves (circles in Fig. \ref{fig:exgal} representing individual molecular clouds) match these global dense-gas star formation relations quite well, with a similar scatter as other regions.

The finding that molecular-cloud-sized structures in the CMZ (leaves, cyan circles in Figure \ref{fig:exgal}) seem to match the global SFR vs. dense gas relationship is significant. Several previous works have found a significant lack of sub-structure across the CMZ on $\sim$ 0.1 pc scales \citep{Battersby2020, Hatchfield2020, Lu2019a, Lu2019b}. The emerging paradigm seems to be that CMZ gas which is engaged in the star formation process (such as that contained within individual molecular clouds) forms stars as normal. However, much of the CMZ's dense gas (the CMZ overall data points) is simply not engaged in the star formation process.

%Figure now made in sfr_vs_alpha_plot_megatable.ipynb
\begin{figure*}
\centering
\includegraphics[width=0.9\textwidth]{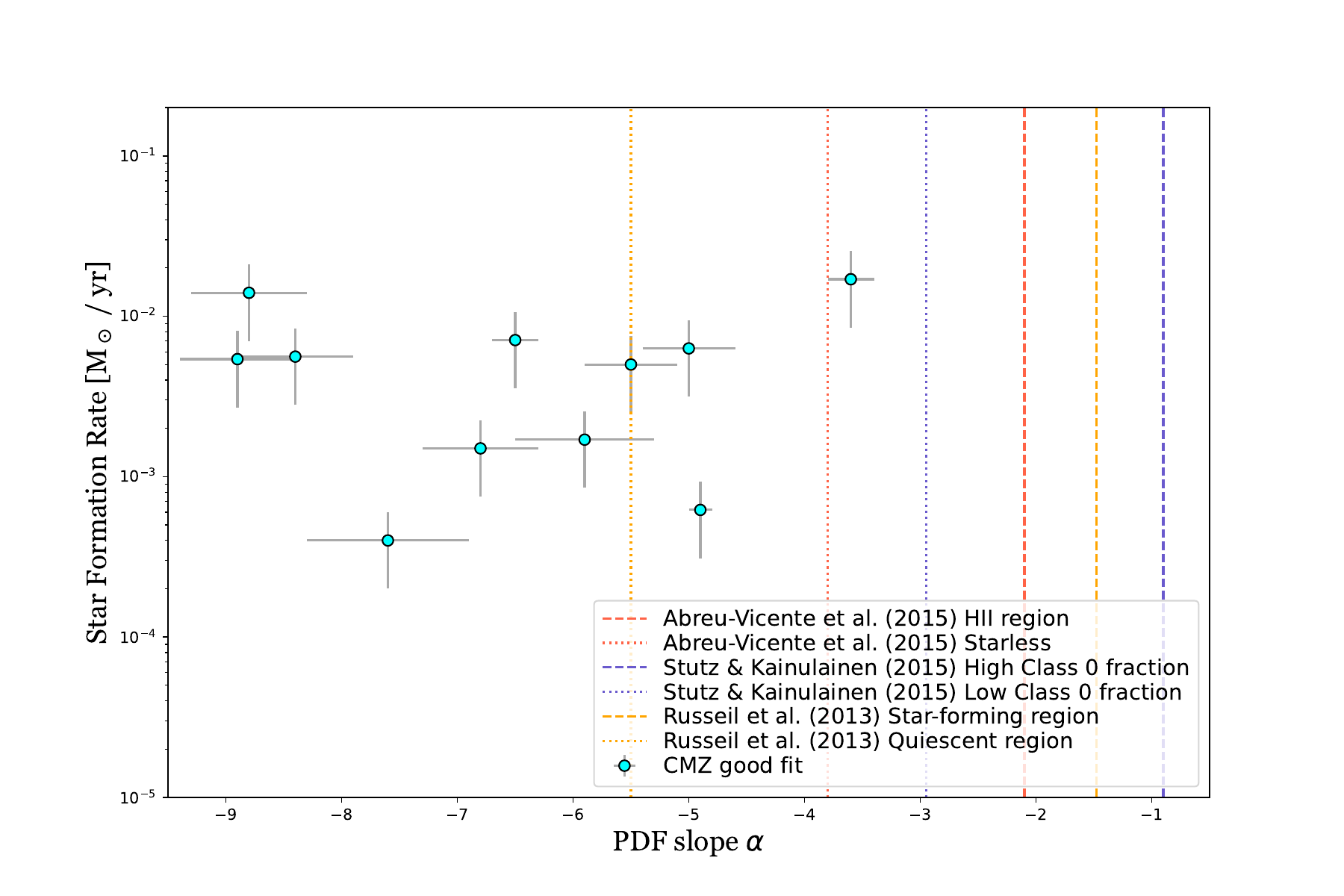}
\caption{A comparison of the N-PDF slope of CMZ structures (in cyan) with their star-formation rate shows at best a marginal trend of steeper N-PDF slope with decreasing SFR, but with very small significance and a number of suspicious outliers.
We plot here the PDF slope $\alpha$ and their errors from Table \ref{tab:powerlaw} (described in further detail in Section \ref{sec:mle}) on the x-axis versus the SFRs from Table \ref{tab:luminosity_properties} (discussed in Section \ref{sec:sfr}) assuming an error of 50\% for each point. 
As described in Section \ref{sec:pdfs} some regions appear to be in good agreement with the overall literature trend of a steeper N-PDF slope indicating a lower SFR, while other regions buck this trend. We plot various literature N-PDF slopes as vertical lines, with dashed lines indicating the more actively star-forming average slopes and dotted lines the more quiescent slopes from these publications. This plot demonstrates also that our slopes tend to be steeper than most reported literature values.}
\label{fig:sfr_pdfslope}
\end{figure*}

\subsection{N-PDFs Power-law fits}
\label{sec:pdfs}

In general, observations of quiescent molecular clouds have N-PDFs that are consistent with a lognormal distribution, indciating that they are dominated by turbulence. Some work, however, \citep[e.g.][]{Alves2017} suggests that molecular cloud N-PDFs are consistent with steep power-law slopes. It is agreed that molecular clouds with signatures of star formation show a shallower power-law tail at high densities, indicating the presence of self-gravity in addition to the turbulent lognormal (or steep power-law) component that dominates at lower densities. These works  demonstrate that more prolific star formation is correlated with a shallower power-law tail \citep{Kainulainen2009, Russeil2013, Abreu-Vicente2015, Stutz2015, Schneider2022}. Both \citet{Russeil2013} and \citet{Stutz2015} analyze the N-PDFs of sub-regions within a high-mass star-forming region (NGC6334 and Orion, respectively) using Herschel data. \citet{Russeil2013} find the shallowest power-law slope in the central star-forming region (p = -1.48) with steeper slopes in the less-evolved regions (from p= -3.14 to p = -5.58). \citet{Stutz2015} find that the regions with the highest Class 0 fractions have the shallowest slopes (down to p = -0.9) and the regions with the lowest Class 0 fraction have the steepest slopes (up to p = -2.95). \citet{Lim2016} study the N-PDF of an extreme Infrared Dark Cloud and find that it is consistent with a lognormal, with some tentative evidence for a deviation to a power-law at high densities. They find similar results using both a near-IR $+$ mid-IR dust extinction technique and a Herschel Far-IR dust emission, and report that the Herschel Galactic Gaussian method \citep[introduced in ][and used in this work]{Battersby2011} to be well-suited for the subtraction of the large-scale Galactic fore-/back-ground. \citet{Abreu-Vicente2015} perform the most comprehensive study of molecular cloud N-PDFs to date, studying 100s of molecular clouds across three orders of magnitude in mass and spanning the evolutionary sequence from starless, to star-forming, to {\sc Hii} regions using ATLASGAL data. They find that starless regions have N-PDFs consistent with a lognormal distribution, while star-forming regions show a mix of lognormal at low-densities and a power-law tail at high densities, with an integrated average power-law slope of p = -3.8. The {\sc Hii} region population is dominated by a shallower power-law, with an integrated average slope of p = -2.1. Similar results are reported by \citet{Schneider2022}. Work by \citet{Chen2018} studies the anatomy of N-PDFs using dendrograms on both observations and simulations and find that individual sub-structures have power-law indices that are different from the N-PDF of the entire region, which may indicate different stages of gravitational collapse. 

The slopes of our N-PDFs are generally quite steep compared with literature values (typically $-6 < \alpha < -1$) and range from $\alpha = -2.5$ at the shallowest (SgrB2 overall region) to $\alpha = -8.9$ (the 1.1\deg~West cloud, ID: 39) at the steepest, as shown in Figure \ref{fig:sfr_pdfslope}. A steep power law can indicate non-star-forming turbulent gas, that could also have been fit with a lognormal. If these regions reflect the turbulent, chaotic gas rather than star-forming gas, then a steep power-law slope is not surprising. Finding strong turbulence even at these high densities is not surprising in this part of the Galaxy \citep[e.g.][]{Henshaw2023, Henshaw2016a, Ginsburg2016, Shetty2012}. 

We present the power-law slopes in Table \ref{tab:powerlaw}. These are plotted against the best SFR estimate (from Table \ref{tab:luminosity_properties}) in Figure \ref{fig:sfr_pdfslope}. After all of the PDF exclusions based on quality of fit (see Section \ref{sec:mle}) and one missing SFR (ID: 29, region was a branch but was too small for the IR-based SFR method, see Section \ref{sec:sfr} for details), we are left with eleven data points. 

As described earlier in this section, typically, a shallower power-law tail is thought to be associated with more vigorous star formation. We see this trend in our overall region fits. The Sgr B2 region is considered one of the most actively star-forming regions in the Galaxy and has the shallowest power law overall ($\alpha = -2.5 \pm 0.1$), the outer 100 pc overall region is largely considered non-star-forming and has the steepest power law overall ($\alpha = -7.3 \pm 0.2$), while the inner 100 pc overall region is in the middle ($\alpha = -3.3 \pm 0.1$). 
However, the results are not uniformly in agreement on the individual cloud scale. 
Figure \ref{fig:sfr_pdfslope} shows a correlation in the majority of the regions (7 of the 11) with three outliers significantly above, and one outlier significantly below this trend in their SFR. The three high outliers are 1.1\deg~west (ID: 39), the bubble (ID: 21) \citep{Nonhebel2024}, and SgrB2 extended (ID: 30) and the one low outlier is the 1.6\deg~complex (ID: 23).

Column density maps only tell us part of the story of the CMZ, but N-PDF slopes allow us to connect this extreme region of the Galaxy with our solar neighborhood, simulations, and other regions. The conclusion that the PDF slope in the CMZ is generally steeper than others reported in the literature may be due to the fact that many of our structures sample large physical areas, with a high fraction of dense gas that is not involved in the star formation process \citep{Battersby2020}. 
We note that the majority of the PDFs are complex and may not be well-described by a single power-law \cite[compare with the multiple power-laws reported by][]{schneider2015, Schneider2022}. We have selectively chosen a handful (eleven) regions, plus the three overall CMZ regions, that appear to be well-fit by this metric for our analysis and to be able to compare with other regions.

\section{Conclusions}
\label{sec:conclusions}

The CMZ is the hub at the center of the Milky Way Galaxy that connects gas flowing from Galactic scales with the central nucleus. We present a hierarchical structure catalog of the CMZ and measurements of the physical and kinematic properties, as well as SFR estimates of each structure. We compare the properties derived with other regions in the Milky Way and external galaxies. As part of this work we release the full dendrogram hierarchical structure and the full catalog in the form of a machine-readable table that includes all of the columns from Tables \ref{tab:general_properties}-\ref{tab:powerlaw} in one coherent framework. Below, we summarize key conclusions drawn from this work: 

\begin{itemize}

\item Using the column density map from Paper I (Battersby et al., submitted), we develop a multi-scale catalog of dense gas structures in the CMZ using {\sc astrodendro}. We identify 11 levels of hierarchical structure, with a total of 57 branches and leaves, with structures on the scale of the entire CMZ down to  individual molecular clouds. 

\item We compute and report the physical properties of each structure in our dendrogram hierarchy. Using Herschel data products from Paper I, we report the area, central coordinates, column densities, dust temperatures, masses, radii, and free-fall times of each structure. 

\item We calculate kinematic properties of each structure using integrated spectra toward each structure with the \citet{Jones2012} MOPRA 3mm data. We compute the kinematic properties using simple moment analysis in three molecular line transitions: HNCO 4(0,4) - 3(0,3), HCN 1-0 (F=2-1), and HC$_3$N 10-9.

\item We integrate the Paper I modified blackbody fits to compute the ``cool" far-IR luminosity of each structure (incorporating wavelengths from 160-500 \micron) and combine with a ``warm" luminosity estimate (adding in wavelengths 5.8-24 \micron~from Spitzer)  from \citet{Barnes2017} to derive a total IR luminosity for each structure. 

\item For the branches in our hierarchical tree with a total area greater than 1000 pc$^2$, we use the \citet{Kennicutt1998} scaling relationship, to convert total IR luminosities to SFR estimates.

\item For leaves in our catalog, which mostly represent individual molecular clouds, we use two methods to estimate SFRs. Where available, we use the SFRs from \citet{Hatchfield2024} using the CMZoom survey \citep{Battersby2020}. In the remainder of leaves we develop a new `free-fall' SFR estimation method based on the work from \citet{Barnes2017} which uses the total luminosity of a structure to infer the embedded stellar mass, then the free-fall time as the timescale for star formation. This method agrees well with the \citet{Hatchfield2024} SFRs in overlapping regions. 

\item We compare our structure sizes, linewidths, and gas surface densities with other molecular clouds in the Milky Way \citep{Heyer2009}, NGC300 \citep{Faesi2018}, and the ten most nearby PHANGS-ALMA galaxies from \citet{Rosolowsky2021}. Similar to previous work, we find that structures in the CMZ are either gravitationally unbound or could be consistent with virial equilibrium if they are in the presence of a high external pressure (P$_e$/k $> 10^{7-9}$ K cm$^{-3}$). Using simplistic physical assumptions, we estimate the external pressure for each leaf in our catalog, based on the thermal, turbulent, and gravitational pressure of its surrounding parent branch. We find that the majority of our leaves are consistent with being in pressure-bounded equilibrium.

\item We place the gas densities and SFRs of our CMZ structures in the context of a variety of systems, from Milky Way clouds to high-redshift galaxies, by comparing their positions on standard SFR vs.\ gas surface density and dense gas mass plots. It has already been reported \citep[e.g.][]{Longmore2012, Immer2012, Henshaw2023} that the CMZ overall is in modest agreement with the Schmidt-Kennicutt star formation relation and falls well below the Gao-Solomon dense gas star formation relation. However, we find that individual molecular-cloud-sized CMZ structures align well with other Milky Way and extragalactic regions and are in agreement with these star formation relationships.

\item Finally, we perform simple power-law fits to the N-PDF of three main CMZ overall regions (defined by natural boundaries in the column density map to be the Outer 100 pc region, Sgr B2 extended, and the Inner 100 pc region) as well as to a handful of structures in the catalog.
We find that our power-law slopes are steeper than those in the literature ($-9 <\alpha < -2$ in this work compared with $-6 <\alpha < -1$ in the literature).
Most of our regions hint at a general trend of steeper N-PDF slopes for lower SFR.

\end{itemize}

\acknowledgments
C.\ Battersby  gratefully  acknowledges  funding  from  National  Science  Foundation (NSF) under  Award  Nos. 1816715, 2108938, 2206510, and CAREER 2145689, as well as from the National Aeronautics and Space Administration through the Astrophysics Data Analysis Program under Award ``3-D MC: Mapping Circumnuclear Molecular Clouds from X-ray to Radio,” Grant No. 80NSSC22K1125.
D. Walker and D. Lipman gratefully acknowledge funding from the NSF under Award No. 1816715 and D. Lipman also acknowledges funding from the National Science Foundation under Award Nos. 2108938, and CAREER 2145689, as well as from the NASA Connecticut Space Grant Consortium under PTE Federal Award No: 80NSSC20M0129.
A. Ginsburg acknowledges funding from NSF awards AST 2008101 and 2206511 and CAREER 2142300.
J.\ D.\ Henshaw gratefully acknowledges financial support from the Royal Society (University Research Fellowship; URF/R1/221620). 
R.\ S.\ Klessen and S.\ C.\ O.\ Glover acknowledge financial support from the European Research Council via the ERC Synergy Grant ``ECOGAL'' (project ID 855130),  from the German Excellence Strategy via the Heidelberg Cluster of Excellence (EXC 2181 - 390900948) ``STRUCTURES'', and from the German Ministry for Economic Affairs and Climate Action in project ``MAINN'' (funding ID 50OO2206). 
E.A.C.\ Mills  gratefully  acknowledges  funding  from the National  Science  Foundation  under  Award  Nos. 1813765, 2115428, 2206509, and CAREER 2339670.
M. C. Sormani acknowledges financial support from the European Research Council under the ERC Starting Grant ``GalFlow'' (grant 101116226).
Q. Zhang acknowledges the support from the National Science Foundation under Award No. 2206512.
Herschel is an ESA space observatory with science instruments provided by European-led Principal Investigator consortia and with important participation from NASA.
COOL Research DAO is a Decentralized Autonomous Organization supporting research in astrophysics aimed at uncovering our cosmic origins.

\software{This work made use of Astropy:\footnote{http://www.astropy.org} a community-developed core Python package and an ecosystem of tools and resources for astronomy \citep{astropy:2013, astropy:2018, astropy:2022}.
This research made use of astrodendro, a Python package to compute dendrograms of Astronomical data (\href{http://www.dendrograms.org/}{http://www.dendrograms.org/}) as well as SAO Image ds9 \citep{Joye2003}.
This research made use of Montage. It is funded by the National Science Foundation under Grant Number ACI-1440620, and was previously funded by the National Aeronautics and Space Administration's Earth Science Technology Office, Computation Technologies Project, under Cooperative Agreement Number NCC5-626 between NASA and the California Institute of Technology.
This research has made use of NASA's Astrophysics Data System Bibliographic Services.}

\clearpage

\bibliography{refs_3dcmz.bib}{}

\end{document}